\documentclass[journal,english,twocolumn]{IEEEtran}
\pdfoutput=1

\usepackage[T1]{fontenc}
\usepackage{amsmath}
\usepackage{graphicx}
\usepackage{amssymb}
\usepackage{subfigure}

\usepackage{gensymb}

\usepackage[dvipsnames]{xcolor}

\makeatletter

\usepackage{cite}

\usepackage[caption=false,font=footnotesize]{subfig}

\makeatother

\usepackage{tikz}
\newcommand\copyrighttext{%
  \footnotesize \textcopyright 2015 IEEE. Personal use of this material is permitted.
  Permission from IEEE must be obtained for all other uses, in any current or future 
  media, including reprinting/republishing this material for advertising or promotional 
  purposes, creating new collective works, for resale or redistribution to servers or 
  lists, or reuse of any copyrighted component of this work in other works. }
\newcommand\copyrightnotice{%
\begin{tikzpicture}[remember picture,overlay]
\node[anchor=south,yshift=0pt] at (current page.south) {\fbox{\parbox{\dimexpr\textwidth-\fboxsep-\fboxrule\relax}{\copyrighttext}}};
\end{tikzpicture}%
}

\usepackage{babel}

\begin{document}
\bstctlcite{IEEEexample:BSTcontrol}

\title{Characterization of a Low-Frequency Radio Astronomy Prototype Array in Western Australia}

\author{A. T. Sutinjo, T. M. Colegate, R. B. Wayth, P. J. Hall, E. de Lera Acedo, T. Booler, A. J. Faulkner, L. Feng, N. Hurley-Walker, B. Juswardy, S. K. Padhi, N. Razavi-Ghods, M. Sokolowski,  S. J. Tingay, \and J. G. Bij de Vaate

}

\maketitle
\copyrightnotice

\begin{abstract}
We report characterization results for an engineering prototype of a next-generation low-frequency radio astronomy array. This prototype, which we refer to as the Aperture Array Verification System 0.5 (AAVS0.5), is a sparse pseudo-random array of 16 log-periodic antennas designed for 70--450~MHz.   It is co-located with the Murchison Widefield Array (MWA) at the Murchison Radioastronomy Observatory (MRO) near the Australian Square Kilometre Array (SKA) core site. We characterize the AAVS0.5 using two methods: in-situ radio interferometry with astronomical sources and an engineering approach based on detailed full-wave simulation. In-situ measurement of the small prototype array is challenging due to the dominance of the Galactic noise and the relatively weaker calibration sources easily accessible in the southern sky. The MWA, with its 128 ``tiles'' and up to 3~km baselines, enabled in-situ measurement via radio interferometry. We present  array sensitivity and beam pattern characterization results and compare to detailed full-wave simulation. We discuss areas where differences between the two methods exist and offer possibilities for improvement. Our work demonstrates the value of the dual astronomy--simulation approach in upcoming SKA design work.

\end{abstract}


\thispagestyle{empty}

\section{Introduction}
\label{sec:intro}

\begin{figure}[htb]
 \begin{center}
 \includegraphics[width=3.5in]{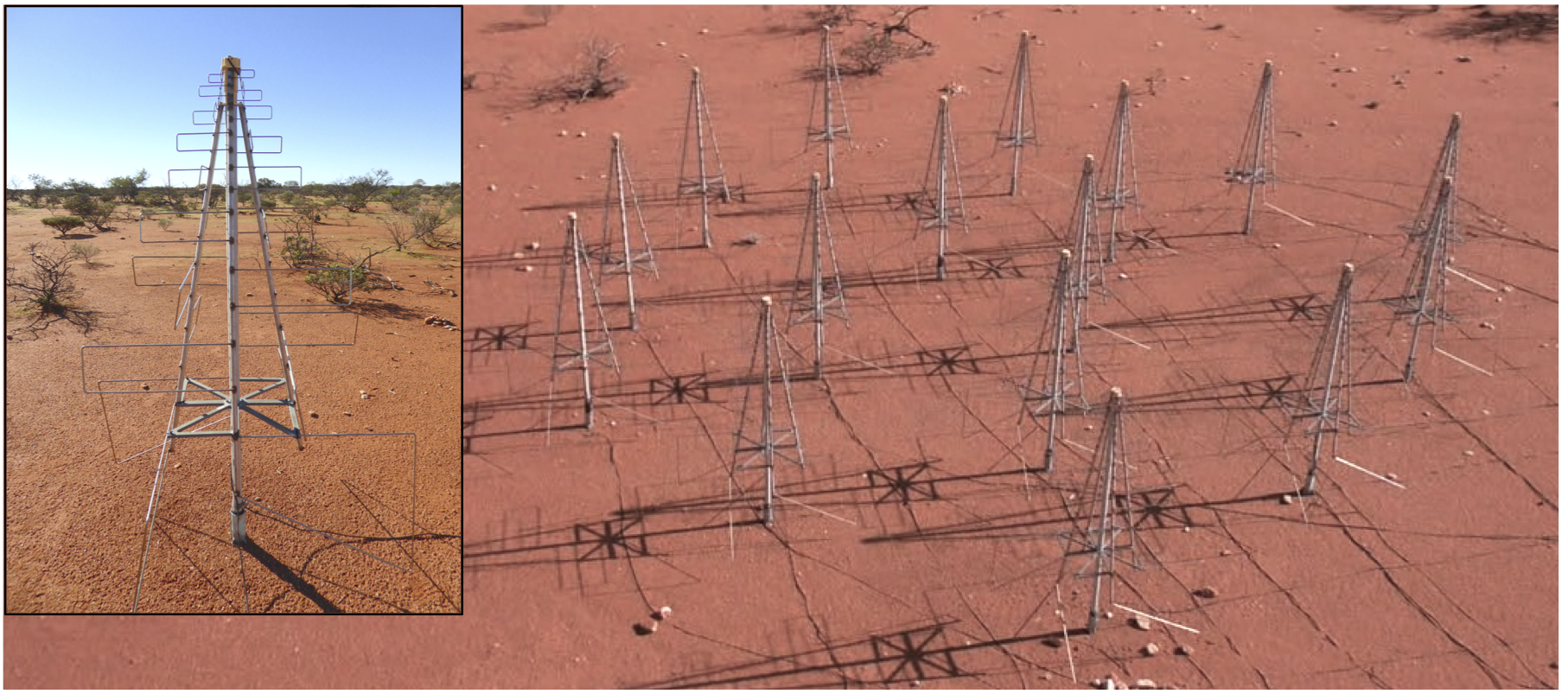}
 \caption{The AAVS0.5 is an array of 16 dual-polarized SKALA antennas (inset) pseudo-randomly placed in an 8~m diameter circle. In current implementation, the antennas are mounted over soil without a groundplane.}
 \label{fig:jan2014}
 \end{center}
 \end{figure}

In the current pre-construction phase of the Square Kilometre Array (SKA) radio telescope~\cite{HalSch08,5136190, 6051197}, verification systems are essential in demonstrating that candidate designs meet functionality, cost target and site requirements. To demonstrate potential telescope and sub-system designs, a series of Aperture\footnote[1]{The term ``aperture'' array in Radio Astronomy context is not related to aperture antennas in antenna engineering sense. It simply refers to a phased array, as opposed to a parabolic dish, which acts as an aperture that converts power density from celestial sources to received power.} Array Verification Systems (AAVSs) are being constructed. The AAVS0~\cite{6328681, EloyEuCap2012}  was the first prototype constructed in Lord's Bridge (UK). The AAVS0.5, shown in Fig.~\ref{fig:jan2014}, is the first verification system in Australia\footnote[2]{constructed by the International Centre for Radio Astronomy Research (ICRAR), University of Cambridge and the Netherlands Institute for Radio Astronomy (ASTRON)}~\cite{AAVS05_ICEAA13}. In the near future, the AAVS0.5 will be succeeded by the AAVS1, a larger prototype\footnote[3]{to be constructed by the SKA Aperture Array Design and Construction Consortium http://www.skatelescope.org/lfaa/} having at least one phased array of 256 antennas corresponding to a full SKA low-frequency aperture array station.
 
In-situ measurement of a radio astronomy array is a necessary step in its design verification. It entails understanding the array behavior in the intended physical environment and its response to celestial radio emission. For antenna array characterization, bright compact cosmic radio sources, with some caveats, are convenient tools, as their positions on the celestial sphere are well known and they emit broadband noise~\cite{4315019}, enabling wide band measurement of the antenna and receiver under test (AUT)~\cite{4315019, 1140521}.

However, at low frequencies such sources are not numerous in the southern sky. Furthermore, the Galactic noise, which is a greatly extended rather than compact source, is dominant~\cite{Ellingson_TAP2011, LWA1_6420880, Wijnholds_TAP2011}. 
As we  will discuss in detail, we overcome these challenges through low-frequency radio interferometry, which for the AAVS0.5 is performed in conjunction with the Murchison Widefield Array (MWA)~\cite{2013PASA...30....7T, Lonsdale_2009}. The MWA is one of several operating low-frequency radio interferometers, including LOFAR~\cite{2013A&A...556A...2V} and PAPER~\cite{1538-3881-139-4-1468}.

This paper is organized as follows. Sec.~\ref{sec:AAVS05desc} provides a description of the AAVS0.5 array. Full-wave simulation of the array is discussed in Sec.~\ref{sec:full-wave}. Sec.~\ref{sec:background} reviews the method of characterization via radio interferometry. Sec.~\ref{sec:meas} reports measurement results and comparison to full-wave simulation. Concluding remarks are given in Sec.~\ref{sec:concl}.

\section{AAVS0.5 Description}
\label{sec:AAVS05desc}
\subsection{Physical Description}
\label{sec:construction}

\begin{figure}[htb]
\begin{center}
\includegraphics[width=2.5in]{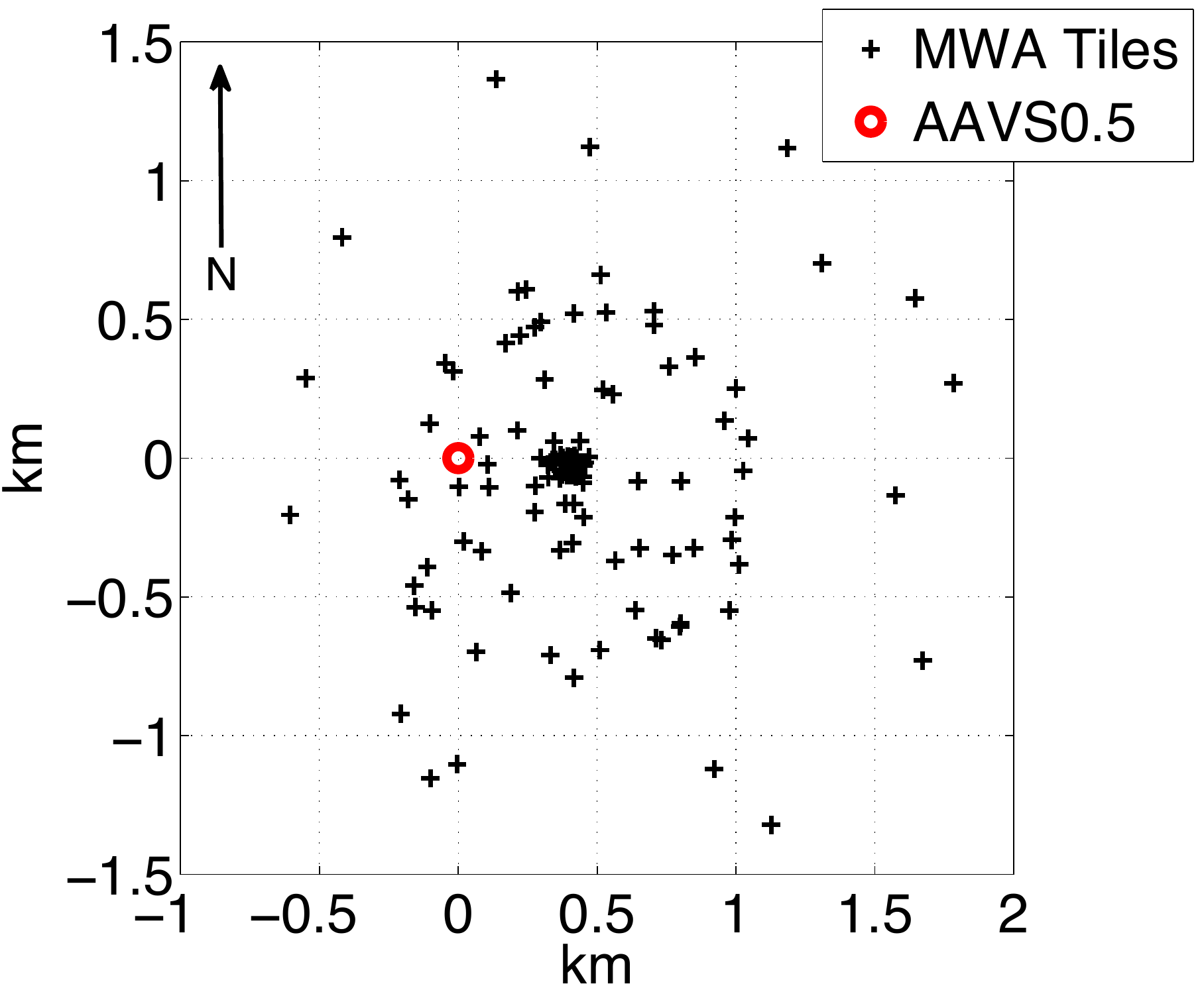}
\caption{A map of the location of the AAVS0.5 (red circle at 0,0) and the MWA telescope tiles (black cross). The center of the AAVS0.5 is located at Latitude/Longitude -26~$^{\circ}$ 42~' 2.59786~''~/~116~$^{\circ}$ 39~' 57.74116~''} \label{fig:AAVS05_mwa}
\end{center}
\end{figure}

Fig.~\ref{fig:AAVS05_mwa} shows the AAVS0.5 location, which is within the extent of the MWA ``tiles'' (each of which is an sub-array of 4$\times$4 bow-tie antennas). This arrangement offers up to 127 MWA tiles at the sufficiently large distances to the AAVS0.5 that are necessary for its characterization (further discussion in Sec.~\ref{sec:background}).

\begin{figure}[htb]
\begin{center}
\includegraphics[width=2in]{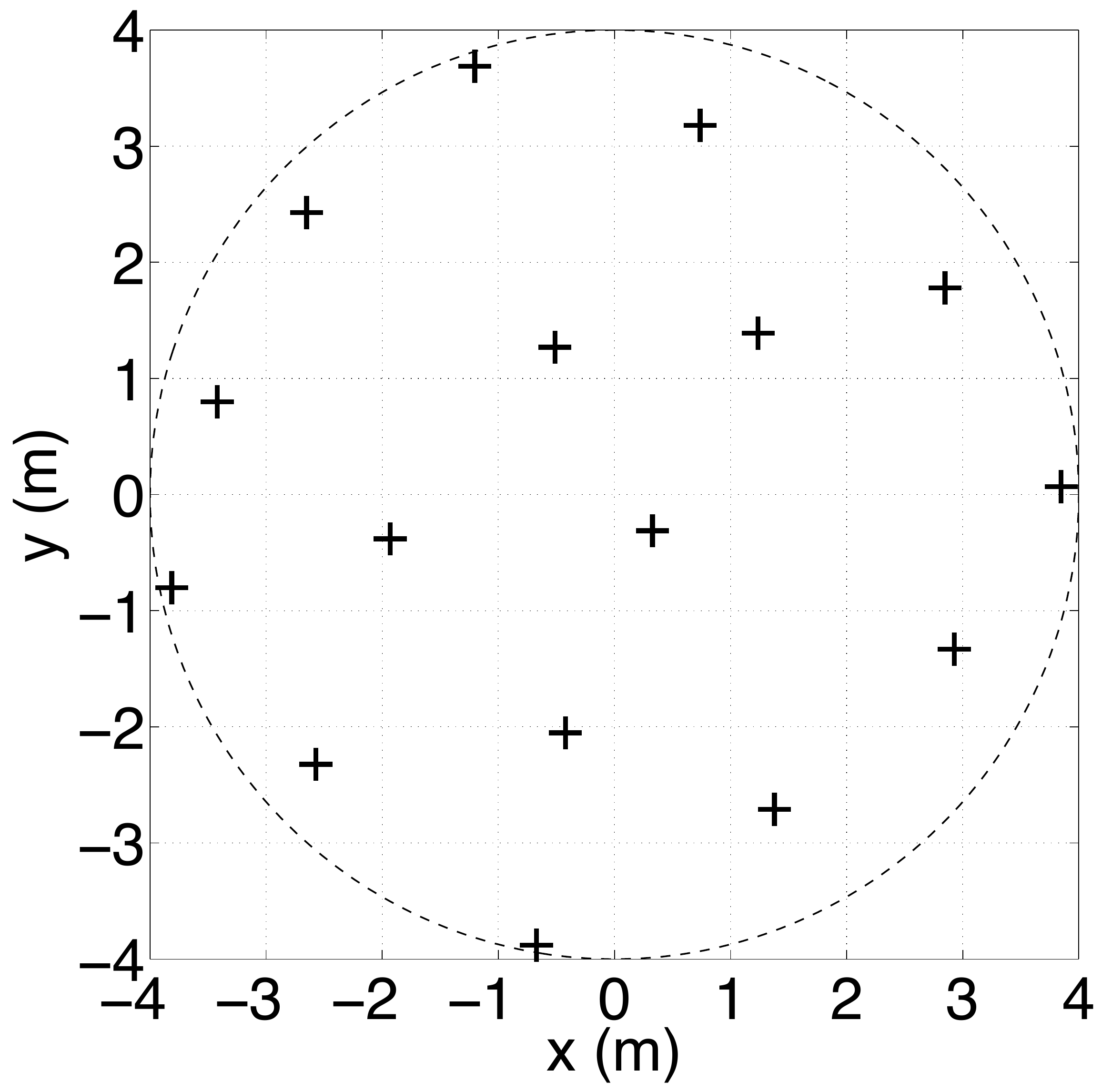}
\caption{The AAVS0.5 as-built points. In the field, y-axis is aligned to North. The as-built points are the locations of the bases of the SKALAs established through professional differential GPS survey post construction. }
\label{fig:as_built}
\end{center}
\end{figure}

The AAVS0.5 consists of 16 dual-polarized log-periodic ``SKALAs'' capable of operation from 50 to 650~MHz~\cite{Eloy_EXAP2015}. Fig.~\ref{fig:as_built} illustrates the pseudo-random positions of the SKALAs located within an 8 m diameter circle. The sparse pseudo-random placement, as opposed to regular spacing, has been selected to prevent the appearance of grating  lobes~\cite{1138220, 2006astro.ph.11160B, 4058261, 5153341} and scan blindess~\cite{EloyICEAA12, 6328681,6632384} over the 50 to 350~MHz SKA bandwidth ($7:1$)~\cite{SKA1base, SKA14-SRS-L1-Rev4}.

At the AAVS0.5 location, the soil type is sandy soil rich in quartz sands and the surface is generally flat to within approximately 2~cm. A layer of granite rock is found underneath the soil which, in the immediate vicinity of the AAVS0.5, is approximately 20 cm in depth on average. The SKALAs are mounted on 60~cm fence posts~\cite{Hicks_LWA_2009} placed in pre-drilled holes.

We constructed the array manually using tools such as a handheld GPS, compass, and measuring tapes. Using a digital inclinometer, the vertical alignment of SKALAs was optimized to better than 1.5$^{\circ}$. Post-construction, the positions and orientations of the SKALAs were professionally surveyed with differential GPS with relative accuracies of the order of a few millimeters. The North-South (N-S) and East-West (E-W) alignment of the SKALAs were determined to be 3$^{\circ}\pm 1.5^{\circ}$ (standard deviation). We find this level of rotation tolerance to be well within acceptable limits for polarization performance~\cite{Sutinjo_TAP_TBP2014,6632385}. The array construction process provided us with valuable first-hand experience of actual site conditions and achievable tolerances. 

\subsection{Signal path and processing system}
\label{sec:signal}
\begin{figure}[htb]
\begin{center}
\includegraphics[width=3.5in]{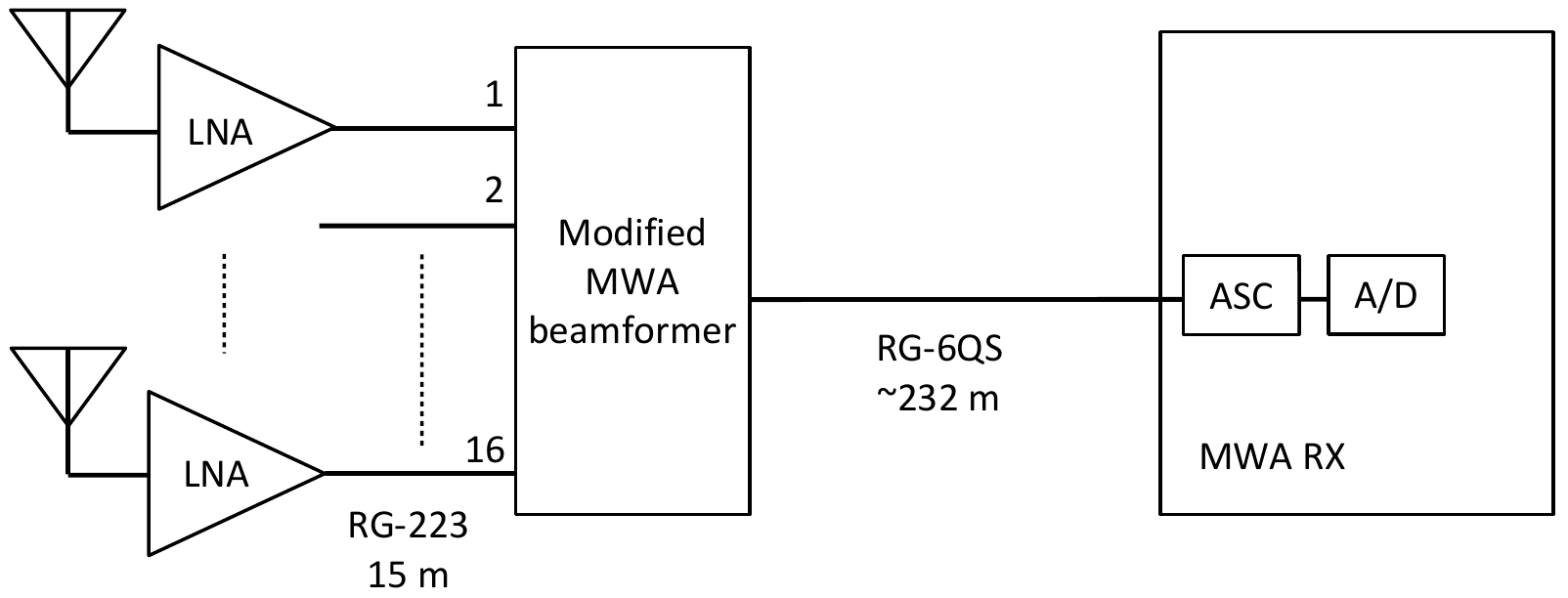}
\caption{The AAVS0.5 RF system block diagram for one polarization of the SKALA. ASC stands for ``analog signal conditioning''. 
RG-223 cables are phased matched solid polyethylene (PE) 50~$\Omega$ coaxial cables. RG-6 cable is 75~$\Omega$ PE coaxial cable provides power and communications to the MWA B/F.} 
\label{fig:AAVS_RX}
\end{center}
\end{figure}

The AAVS0.5 relies on the MWA infrastructure for power, RF signal chain and data processing back-end~\cite{2013PASA...30....7T}. The key components of the signal path are shown in Fig.~\ref{fig:AAVS_RX} and described below.  
\begin{itemize}
\item Low-noise amplifier\footnote[4]{provided by Cambridge Consultants (CC) and the University of Cambridge~\cite{Eloy_EXAP2015}} (LNA): nominal noise temperature and gain at 180~MHz are 33~K and 43~dB respectively. The LNA is mounted at the feed point of each SKALA and powered via the coax from the MWA beamformer (B/F).
\item Modified MWA B/F: modified to accept the LNA, whose gain is higher than a standard MWA LNA. Beamforming is accomplished through 32 analog delay steps, 435~ps each~\cite{2013PASA...30....7T}, allowing the AAVS0.5 to scan to zenith angles (ZA) of
$\sim30^{\circ}$.
\item MWA receiver (RX): conditions and digitizes the signal from the AAVS0.5. 
\end{itemize}
The digitized data are sent to the MWA correlator such that the AAVS0.5 can be used with normal MWA observing~\cite{OrdCro15} (not shown).

\begin{figure}[htb]
\begin{center}
\includegraphics[width=3in]{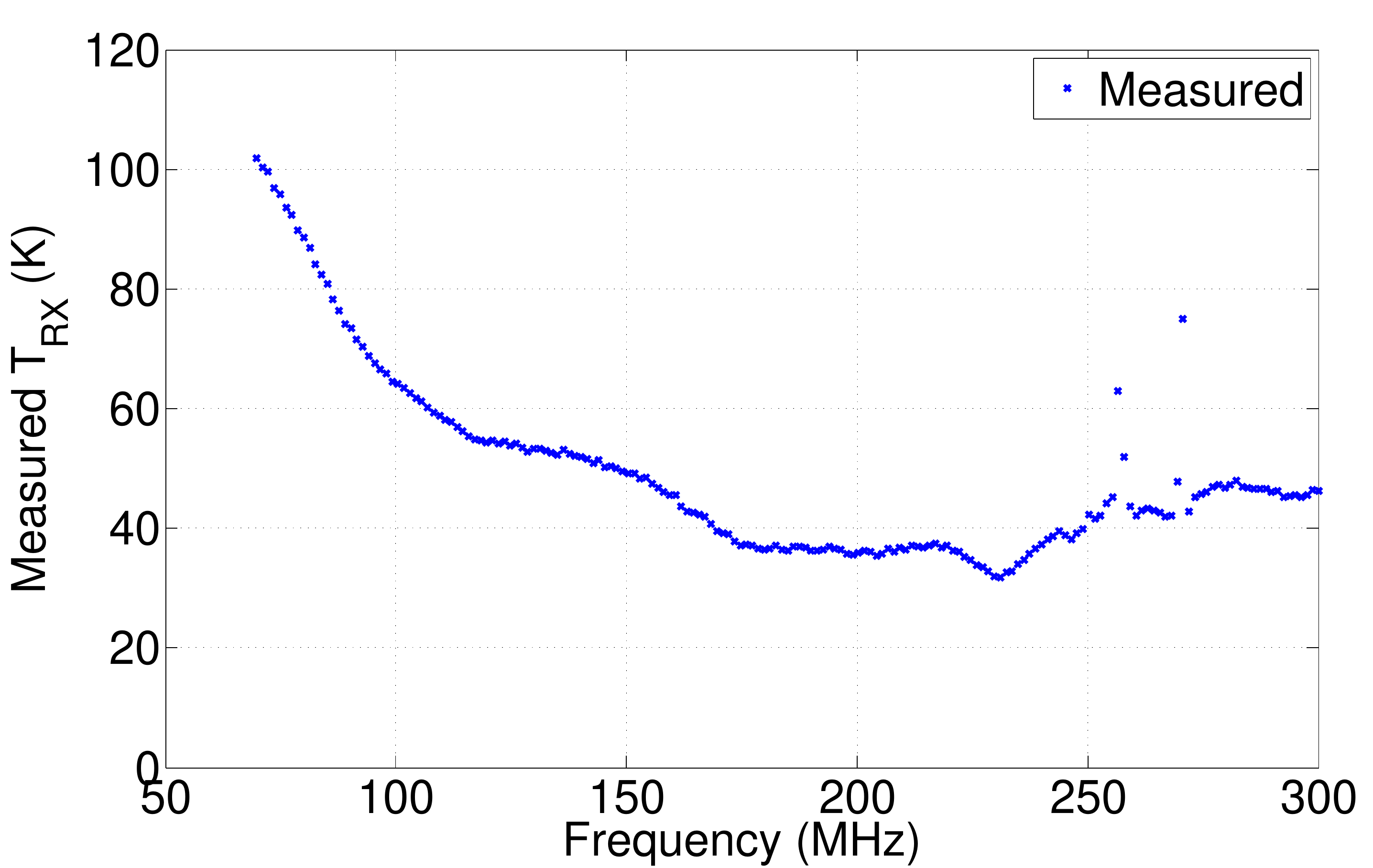}
\caption{Measured receiver noise temperature of the AAVS0.5 RX system.}
\label{fig:TRX}
\end{center}
\end{figure}

Fig.~\ref{fig:TRX} reports the measured receiver noise temperature of the AAVS0.5 system via the RF signal path as described above. 
The measurement was performed on-site using a hot/cold noise source connected to a LNA and then to the rest of the system via a 16-to-1 power divider (de-embedded in post-processing). The outliers between 250--300~MHz are likely due to satellite RFI. 

\section{Full-wave simulation}
\label{sec:full-wave}
An important motivation of the AAVS0.5 characterization is to verify the accuracy of the type of tools available in the design of low-frequency aperture arrays. As the AAVS0.5 does not have a ground plane in this implementation, it is likely to be closely coupled to the environment. As a result, full-wave simulations have become essential in accounting for the effects of soil and cables in the design process. Confirming the accuracy these models is of great interest. 

\begin{figure} [htb]
	\begin{center}
	{\includegraphics[width=3.5in]{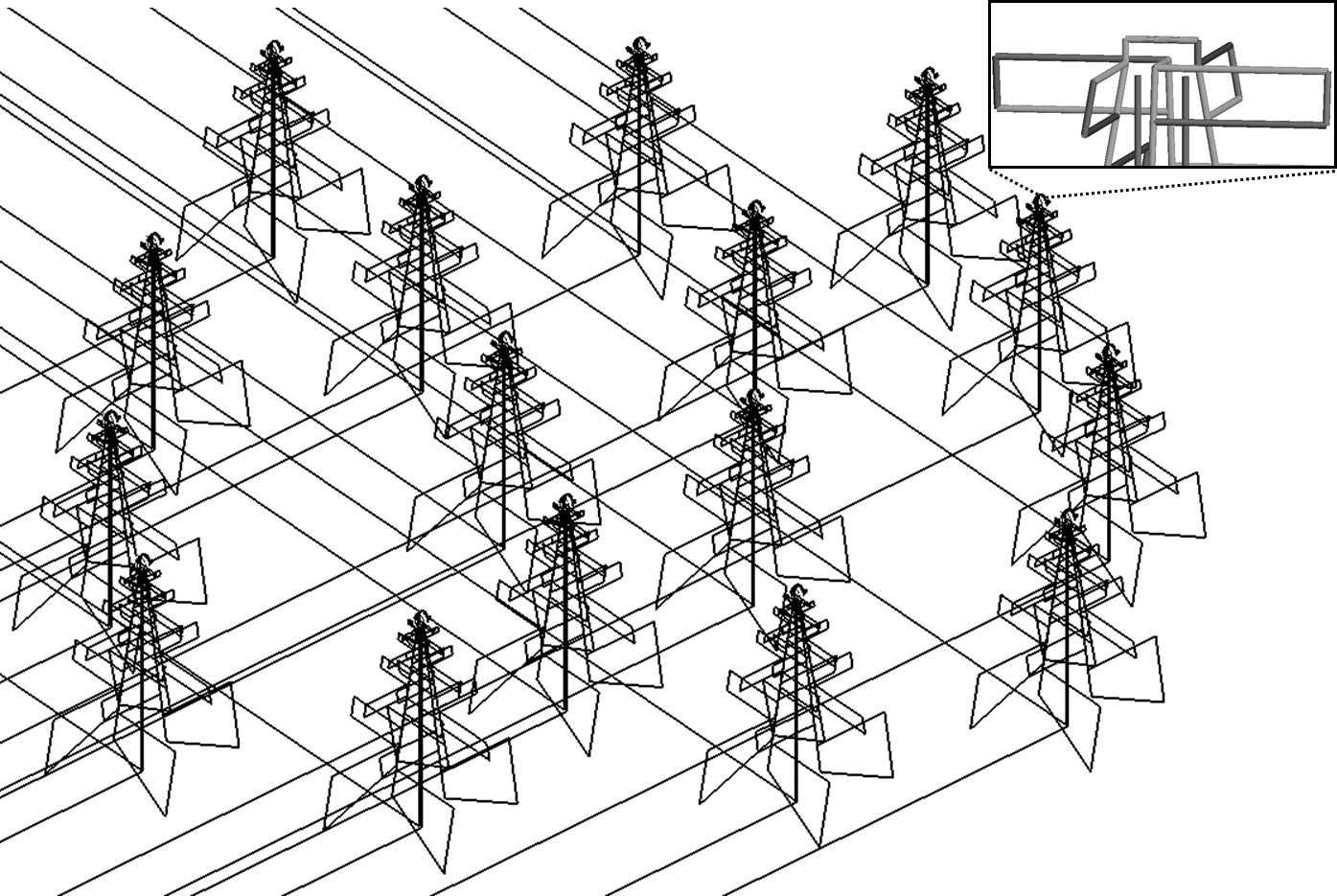}}
	\end{center}
\caption{A screen capture of the 16-element AAVS0.5 antenna array simulated in FEKO. The antenna bases are placed on semi-infinite soil (not shown). The 15~m coaxial cables are modeled as L-shaped wires seen in the figure. The vertical part of the L-shape extends to just below the feed point shown in the inset; the horizontal part is at ground level. Each antenna feed point} is loaded with measured LNA impedance.
\label{fig:FekoModel}
\end{figure}

\begin{figure} [htb]
	\begin{center}
	{\includegraphics[width=\columnwidth]{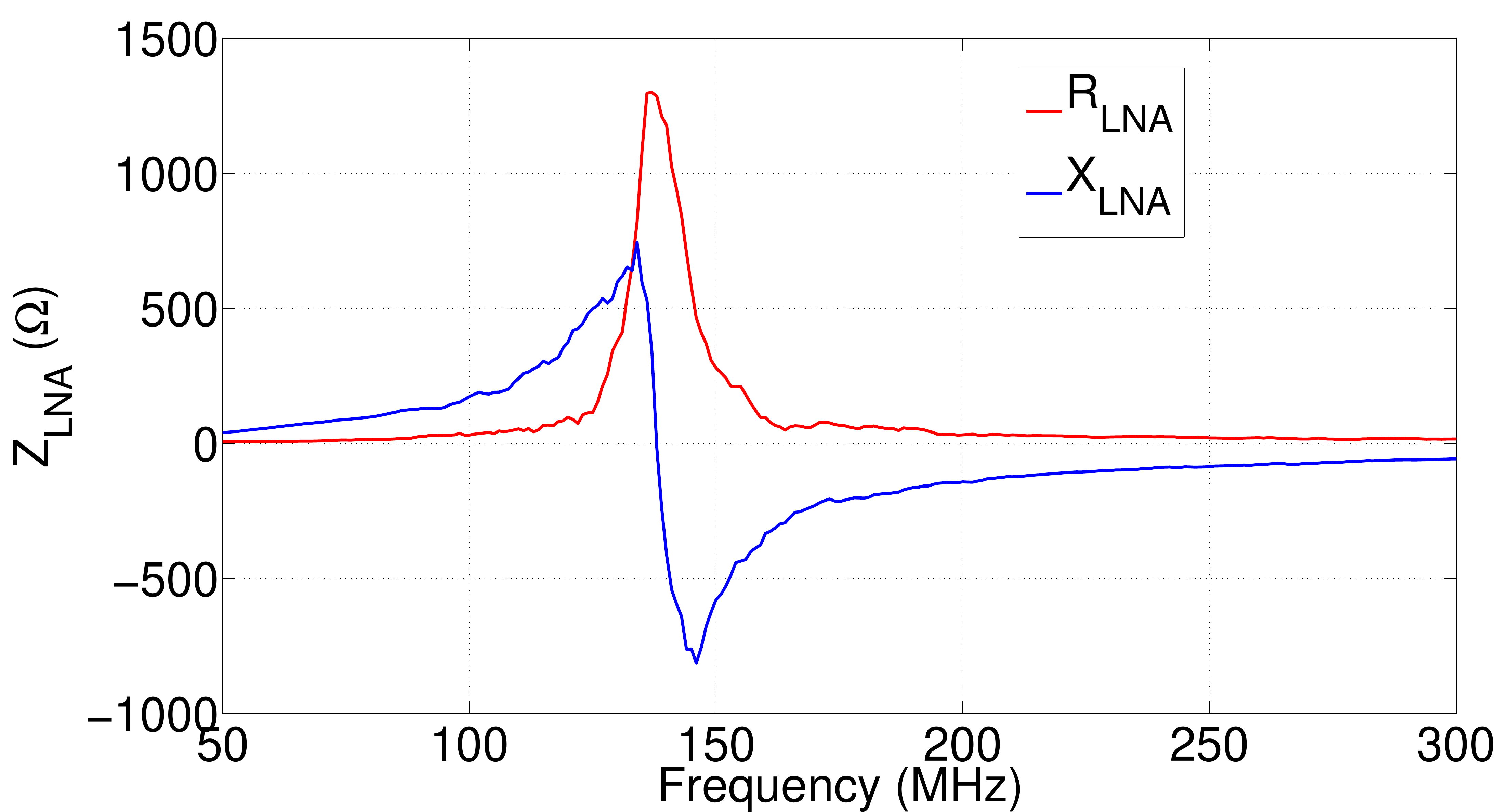}}
	\end{center}
\caption{Measured LNA impedance applied to the antenna ports in FEKO.}
\label{fig:LNAZin}
\end{figure}

Fig.~\ref{fig:FekoModel} depicts a typical AAVS0.5 model in a commercial Method-of-Moments code, FEKO\footnote[5]{www.feko.info}.  The antennas are modeled as a perfect electric conductor (PEC) and the antenna ports are loaded with the measured LNA impedance reported in Fig.~\ref{fig:LNAZin}. The LNA is a balanced (input) to single-ended (output) device with high common mode rejection ratio (CMRR $>\sim35$~dB) in the frequency range of interest. The coaxial cables are simulated as 15~m long wires, 3~mm in diameter. Given the high CMRR, the cables are terminated with open circuits near the feed points as shown in the inset in Fig.~\ref{fig:FekoModel}.
The simulated soil properties are based on measured soil samples from the MRO,
which were characterized experimentally as a function of moisture level,  listed in Tab.~\ref{tab:soil-moist} at spot frequencies.
With the above parameters as input, the beam is simulated for the required pointing direction, where each antenna is given equal amplitude and a relative phase corresponding to the switched-in time delay.

\begin{table} [htb]
	\centering
\caption{Measured relative permittivity $\epsilon_{r}$ and conductivity $\sigma$ (S/m) of MRO soil for different moisture levels.}
\begin{tabular}{c c c c c c c}

\hline
    MHz & \multicolumn{2}{c}{Dry} & \multicolumn{2}{c}{2$\%$} & \multicolumn{2}{c}{10$\%$} \\

    & $\epsilon_{r}$ & $\sigma $ 
     & $\epsilon_{r}$ & $\sigma $
    & $\epsilon_{r}$ & $\sigma $ \\
    	\hline

    50 & 3.9 & 0.0007 & 6.5 & 0.01 & 17.8 & 0.1 \\
   160 & 3.7 & 0.0018 & 5.2 & 0.017 & 14.8 & 0.11 \\
   280 & 3.7 & 0.0022 & 4.8 & 0.02 & 14.4 & 0.11 \\

    	\hline
\end{tabular}
\label{tab:soil-moist}
\end{table}

A common metric for the sensitivity of a radio telescope array is the ratio of antenna effective area to system temperature ($A_{\rm e}/T_{\rm sys}$~$\mathrm{[m^2/K]}$). The antenna\footnote[6]{it is understood that in the context of antenna arrays, the ``antenna'' is an array of antennas} effective area is defined as
\begin{equation}
A_{\rm e}=\frac{\lambda^2\eta\mathcal{D_{\rm max}}}{4\pi}, 
\label{eqn:Ae}
\end{equation}
where $\mathcal{D_{\rm max}}$ is the maximum array directivity for a given pointing direction, and $\eta$ is the radiation efficiency, which includes radiation losses due to the soil as the array is simulated on a semi-infinite lossy dielectric medium in FEKO (Fig.~\ref{fig:FekoModel}).

\begin{figure} [htb]
	\begin{center}
	{\includegraphics[height=2.5in]{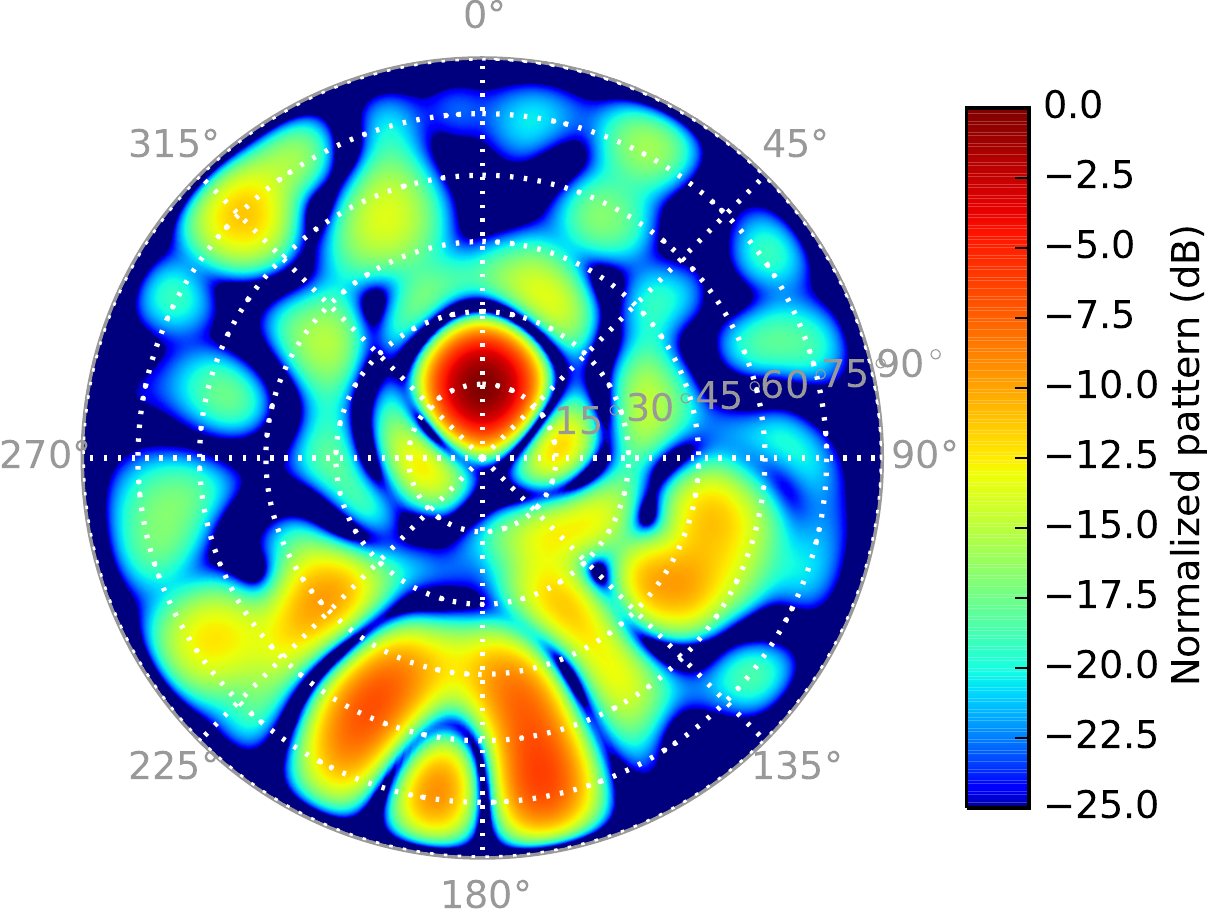}}
	\end{center}
\caption{The FEKO-simulated array normalized beam pattern $\mathcal{D}(\theta,\phi)/D_{\rm max}$for the AAVS0.5 on soil with 2\% moisture at 160~MHz ($\mathcal{D}_{\rm max}=18\,\rm dB$) for the X (East-West) polarization, pointing towards  Az=0\degree, ZA=14.6\degree. The AAVS0.5 half-power beamwidth is $\sim 14^{\circ}$ at this frequency. Radiation efficiency $\eta=0.83$.}
\label{fig:sim-beam}
\end{figure}

\begin{figure} [htb]
	\begin{center}
	{\includegraphics[height=2.5in]{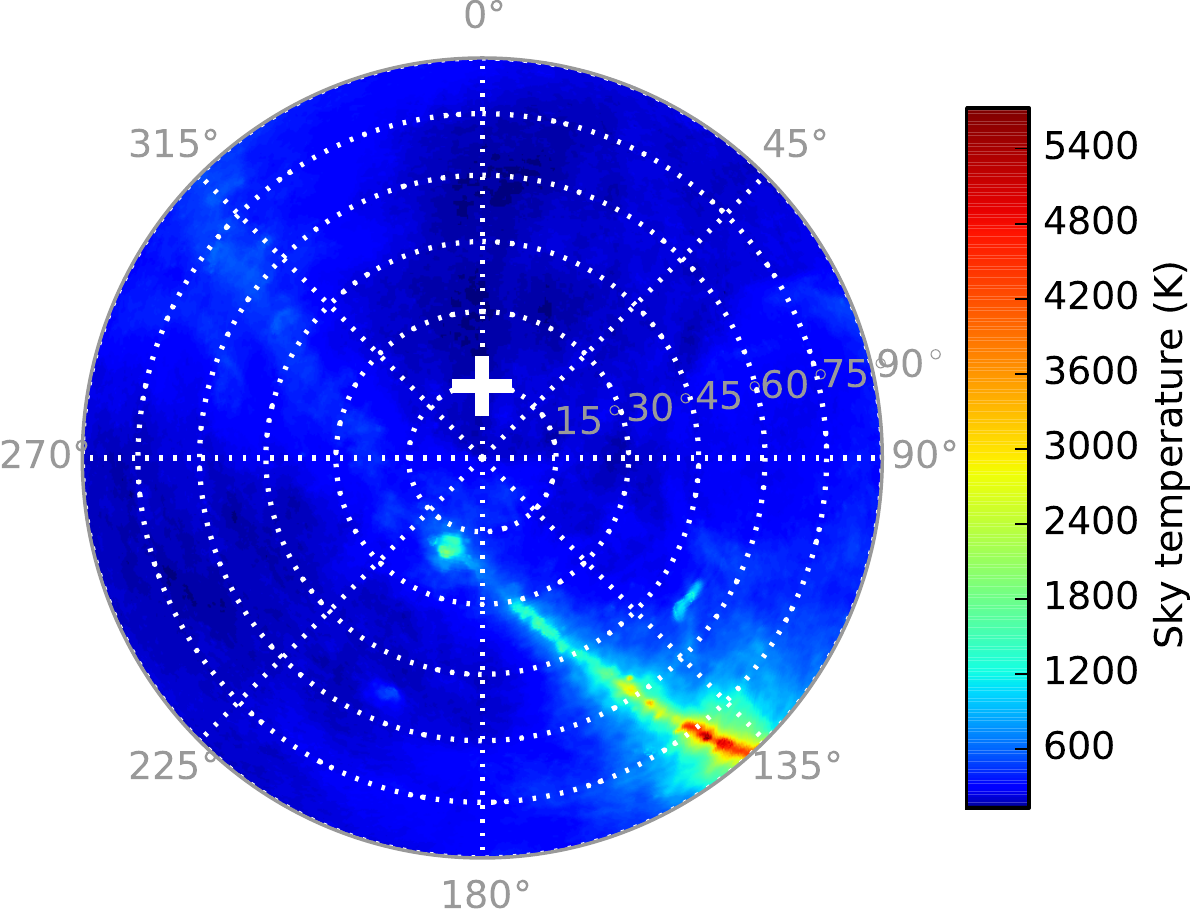}}
	\end{center}
\caption{Sky hemisphere $T_{\rm sky}(\theta,\phi)$ at 160~MHz~\cite{2008MNRAS.388..247D} over the MRO site on
18 January 2015, 1:42 local time (9.27~h local sidereal time). The spatially extended Galactic plane is seen from top-left to bottom-right. Azimuth angle is clockwise from North towards East.
The white plus indicates the location of the calibrator source Hydra A.}
\label{fig:sim-sky}
\end{figure}

The array system temperature is given by:
\begin{equation}
T_{{\rm sys}}=\eta T_{{\rm a}}+(1-\eta)T_{0}+T_{\rm RX},
\label{eqn:Tsys}
\end{equation}
where $T_{\rm 0}$ is the ambient temperature in K, $T_{\rm RX}$ is the measured receiver noise temperature (Fig.~\ref{fig:TRX}) and antenna temperature is
\begin{eqnarray}
T_{\rm a}=\frac{1}{4\pi}\int\mathcal{D}(\theta,\phi)T_{\rm sky}(\theta,\phi)d\Omega,
\label{eqn:Ta}
\end{eqnarray}
where $\mathcal{D}(\theta,\phi)$ is the array directivity pattern (e.g., $\mathcal{D}(\theta,\phi)/D_{\rm max}$ in Fig.~\ref{fig:sim-beam}) in spherical coordinates\footnote[7]{chosen to match MWA and other common practice: $\theta$ (ZA) is the angle away from zenith and $\phi$ (azimuth, Az) is the angle clockwise from North (towards East)}, $T_{\rm sky}(\theta,\phi)$ is the sky under observation
(e.g., Fig.~\ref{fig:sim-sky}) and $d\Omega=\sin\theta d\theta d\phi$ is the infinitesimal element of solid angle.
The first two terms in (\ref{eqn:Tsys}) constitute the total noise temperature seen at the receiver's input due to the antenna; this is referred to as total antenna temperature.
Note that $T_{\rm sky}$  increases exponentially with wavelength such that it is the dominant noise source at low <($\sim200$~MHz) frequencies~\cite{Ellingson_TAP2011, LWA1_6420880, Wijnholds_TAP2011}.
For our simulated A/T, we use sky models based on standard all-sky maps, being relatively low angular resolution observations of the sky at radio frequencies \mbox{~\cite{1982AAS...47....1H, 2008MNRAS.388..247D}} to calculate $T_{\rm a} $ in (\ref{eqn:Ta}).

\section{Characterization of the AAVS0.5 via radio interferometry}
\label{sec:background}

\subsection{Basic Definitions}
\label{sec:defn}
Parameters such as array sensitivity and beam pattern are direction-specific performance measures. 
Direct sensitivity measurement usually involves pointing on and off a bright, compact astronomical source to measure array response in a particular direction~\cite{4315019, LWA1_6420880}. This approach is practicable for an AUT with sufficient sensitivity to detect that object in the presence of system noise. 

In this context, a convenient measure of sensitivity is System Equivalent Flux Density ($\rm SEFD\propto1/(A/T)$), expressed in units of astronomical source flux density Jansky (Jy), where 1~Jy is $10^{-26}$~Wm$^{-2}$Hz$^{-1}$. An astronomical source of $a$~Jy is ``seen'' by radio telescope having SEFD of $b$~Jy with signal-to-noise ratio of $a/b$. The SEFD is defined as
\begin{eqnarray}
\mathrm{SEFD}=2k\frac{T_{\rm sys}}{A_{\rm e}},
\label{eqn:SEFD}
\end{eqnarray}
where $k$ is the Boltzmann constant (using $k=1380$ leads to SEFD in unit of Jy). 

Due to its small size relative to operating wavelength and the high sky noise at 160~MHz, the AAVS0.5 is insufficiently sensitive to discern the difference in total power (on/off source) with integration times of the order of seconds. 
As an illustration, consider the simulated sensitivity of the AAVS0.5 of $\sim0.06~\mathrm{m^{2}/K}$. Using (\ref{eqn:SEFD}), the SEFD of the AAVS0.5 is approximately 46~kJy, which is much greater than easily accessible\footnote[8]{i.e., within the current AAVS0.5 scanning range of ZA$~\sim$30$^{\circ}$.},  ``bright'' ($\sim$0.5~kJy) point-like sources in the southern sky~\cite{LarMil81}.
However, radio interferometry enables us to detect these compact sources and use them for the AAVS0.5 characterization.

\subsection{Brief Review of Radio Interferometry}
\label{sec:review}

A radio interferometer~\mbox{\cite{Kraus_RA_1986, tms01, taylor_synthesis_1999}} may be thought of as a spatial filter that is sensitive to a spatial (or, angular) scale inversely proportional to the separation between the two antennas. Since low-frequency sky noise originates mainly from the Milky Way, which is a large extended source (see Fig.~\ref{fig:sim-sky}), a radio interferometer with significant antenna separation will be insensitive to that source while remaining sensitive to compact cosmic calibration sources.

The property of a radio interferometer may be understood by noting the Fourier Transform relationship between the voltage cross-correlation product as a function of distance between antenna pairs (known as ``visibilities'' as a function of ``baseline length'' in radio astronomy) and the flux density of incoming radiation from the sky as a function of direction.
 This may be illustrated for the simplified 1-D case as
\begin{eqnarray}
V_{ij}(\hat{x})=\left\langle v_{i}v^{*}_{j}\right\rangle(\hat{x})=\int I (\hat{k}_{x}) e^{j2\pi\hat{k}_{x} \hat{x}} d\hat{k}_{x} ,
\label{eqn:visibility1D}
\end{eqnarray}
where $V_{ij}(\hat{x})$ is the visibility measurement between antennas $i$ and $j$, $I (\hat{k}_{x})$ is the flux density of the sky noise incident on the array, $\hat{k}_{x}$ is the normalized free-space wavenumber in the $x$-direction and $\hat{x}=x/\lambda$. 

\begin{figure}[htb]
\begin{center}
\includegraphics[width=\columnwidth]{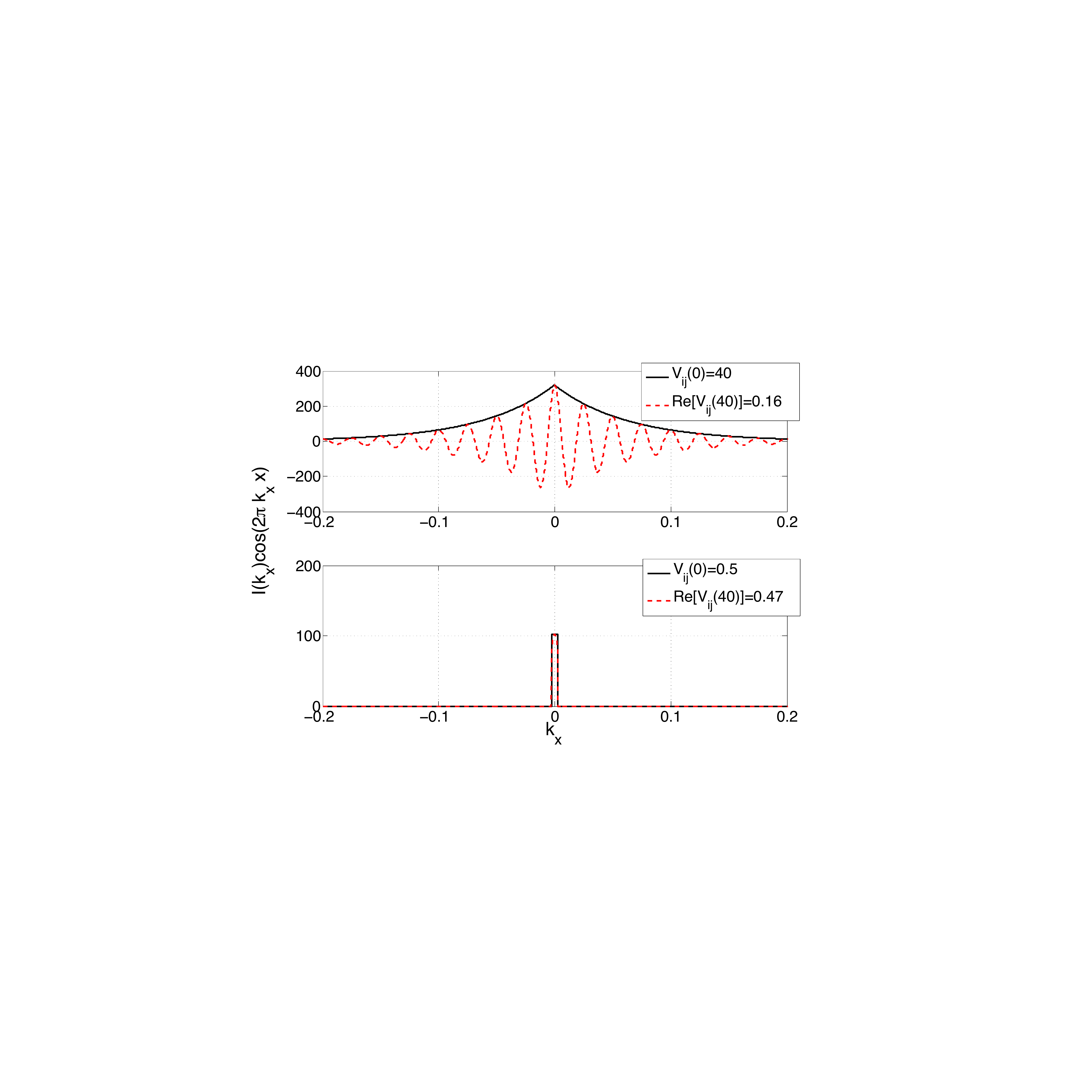}
\caption{A numerical example of (\ref{eqn:visibility1D}), showing the spatial filtering property of an interferometer in 1-D for an extended source (top) and compact source (bottom) for $\hat{x}=0$ and $40$. The legends show the area under each curve.}
\label{fig:fringe}
\end{center}
\end{figure}

Fig.~\ref{fig:fringe} illustrates the spatial filtering property of an interferometer in one dimension described by (\ref{eqn:visibility1D}). For simplicity, consider only the real part of (\ref{eqn:visibility1D}). The black curve in top graph depicts a bright extended source illustrative of Galactic noise as a function of direction. The area under that curve is 40. The black curve in the bottom graph illustrates a compact source with spatial extent $<<1/\hat{x}$, with integrated value 0.5. The red curve in the top graph represents the integrand in (\ref{eqn:visibility1D}) for $\hat{x}=40$ for the extended source; it is a product of the black curve with $\cos(2\pi\hat{k}_{x}40)$. Note that the area under that red curve reduces to 0.16. In contrast, the areas under the curves remain virtually unchanged for the compact source (bottom graph), showing that the interferometer $\hat{x}=40$ is sensitive to the desired compact source but insensitive to the undesired bright extended source.

\begin{figure}[htb]
\begin{center}
\includegraphics[width=1\columnwidth]{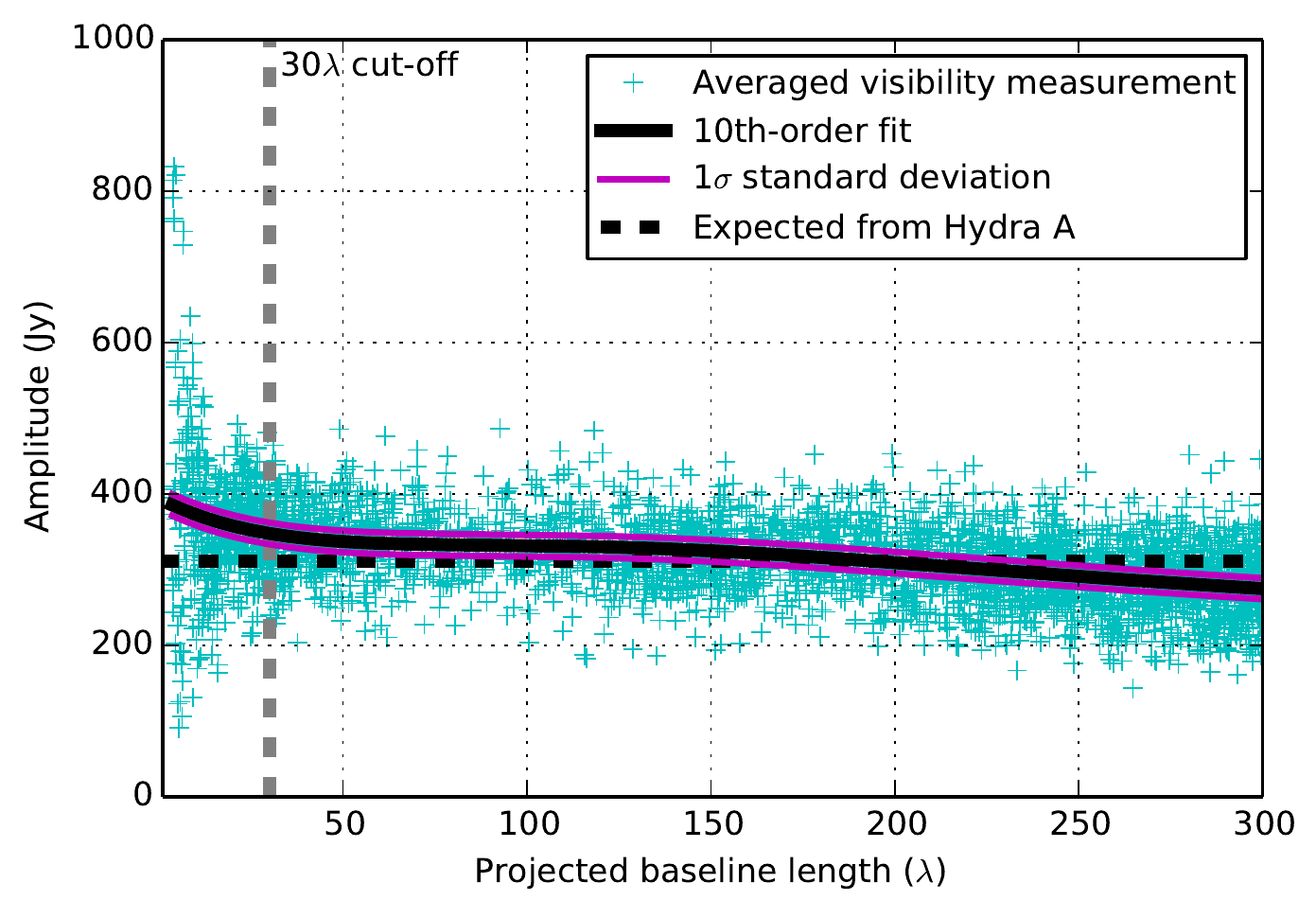}
\caption{Amplitude (Jy) of calibrated MWA visibility measurements as a function of baseline length in wavelengths, projected towards Hydra~A. These data are for a 2-minute snapshot observation of Hydra~A with the MWA and the AAVS0.5 on 18 January 2015. Only a single 40~kHz, X-polarization channel at 160~MHz is shown. 
The black curve is a $10^{\rm th}$-order polynomial fit of the amplitude, and the magenta curves show the $1\sigma$ flux density standard deviation of the mean of each calibrated visibility measurement (see Sec.~\ref{sec:sens_meas}).
The horizontal dashed line shows the expected measurement if the noise contribution was solely from Hydra~A as an idealized point source (310~Jy at 160~MHz).
The vertical dashed line marks the  30$\lambda$ baseline cut-off.
}\label{fig:AAVS0.5_MWA_HydFlux}
\end{center}
\end{figure}

Fig.~\ref{fig:AAVS0.5_MWA_HydFlux} demonstrates this spatial filtering property with an observation corresponding to the sky shown in Fig.~\ref{fig:sim-sky}. To obtain these data, the AAVS0.5 and MWA tiles are electronically pointed to the direction of a Hydra~A (Dec.~$-12\degree06'$, RA~09:18:05 (J2000)), a compact radio galaxy with $\sim0.14^{\circ}$ angular extent~\cite{1538-3881-127-1-48}. Each data point is a visibility measurement of the correlated flux density for a given baseline (more on this in the next Subsection). 

We see in Fig.~\ref{fig:AAVS0.5_MWA_HydFlux} that the known flux density (310~Jy) of Hydra~A dominates the visibilities for baselines of $\sim30-150\lambda$. For baselines shorter than $\sim30\lambda$, the spatially-extended Galactic noise is a significant contributor which causes the mean visibility amplitude to rise. Hence, for the purpose of telescope calibration and the AAVS0.5 characterization, we exclude measurements from baselines of less than 30$\lambda$ (vertical dashed line). The decreasing trend in visibility amplitude for baselines longer than $\sim150\lambda$ is due to partial spatial filtering of Hydra~A itself, the treatment of which will be discussed in Sec.~\ref{sec:meas}.

\subsection{Sensitivity Measurement}
\label{sec:sens_meas}

\begin{figure}[htb]
	\begin{center}
	{\includegraphics[width=3.25in]{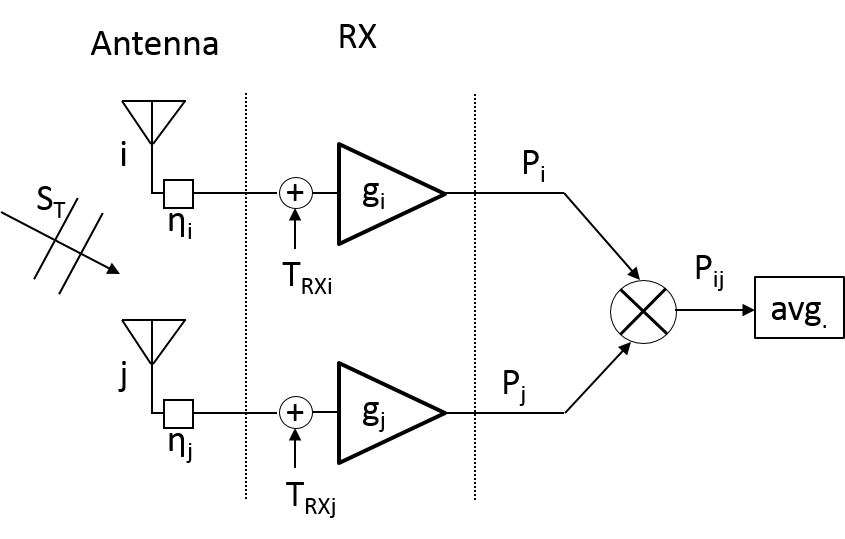}}
	\end{center}
\caption{Signal diagram of a radio interferometer. The system is composed of four parts: antennas, receiver, correlator and averaging.}
\label{fig:Pij}
\end{figure}

In radio interferometry, we infer the sensitivity of an AUT by measuring and normalizing the second-order statistics of the outputs of the interferometer. For brevity, we review only the salient features of this topic and refer interested readers to~\cite{1999ASPC..180..171W} for the full treatment.

Consider the two-antenna interferometer in Fig.~\ref{fig:Pij}. For a single compact source in the telescope's field of view, the expected mean output of the correlator is given by~\cite{1999ASPC..180..171W} 
\begin{eqnarray}
\left\langle P_{ij}\right\rangle = g_{i}g_{j}\sqrt{K_{i}K_{j}}k S_{C} B
\label{eqn:Pij_corr}
\end{eqnarray}
where $K=A_{\rm e}/2k$, $B$ is the receiver noise bandwidth, $k$ is the Boltzmann constant, $g_{i}$ and $g_{j}$ are the voltage gains of receivers $i$ and $j$, respectively; $S_{C}$ is the correlated flux density which is less than or equal to the total flux density $S_{T}$ of the source. If this source is unresolved (not spatially filtered) by the $ij$ baseline then  $S_{C}=S_{T}$. 

We can determine the unknown quantities $g_{i}\sqrt{K_{i}}$ and $g_{j}\sqrt{K_{j}}$ in (\ref{eqn:Pij_corr}) via ``antenna-based gain calibration'': $\left\langle P_{ij}\right\rangle$ is measured (known), $S_{T}$ and $S_{C}$ are assumed known through prior radio astronomical knowledge and the constants $k$ and $B$ are known. 
We determine the antenna-based gains by cross-correlating $N$ antennas which results in $N(N-1)/2$ pairs of antennas (baseline). For each pair~\cite{tms01, 1989ASPC....6...83F, 1989ASPC....6..185C}
\begin{eqnarray}
g_{i}g_{j}\sqrt{K_{i}K_{j}}=\frac{\left\langle P_{ij}\right\rangle}{k S_{C} B}.
\label{eqn:Pij_cal}
\end{eqnarray}
For $N>3$, this problem is overdetermined. Consequently, each $g_{i}\sqrt{K_{i}}$ may be solved via a standard least-squares method \footnote[9]{Knowledge of the antenna based gain allows division of (\ref{eqn:Pij_corr}) by $g_{i}g_{j}\sqrt{K_{i}K_{j}}k B$ such that the measured $\left\langle P_{ij}\right\rangle$ may be expressed in unit Jy. This is referred to as a ``calibrated visibility.''}.

From $g_{i}\sqrt{K_{i}}$ and $g_{j}\sqrt{K_{j}}$, we can obtain AUT sensitivity. The expected variance of the correlator's output is given by~\cite{1999ASPC..180..171W} 
\begin{eqnarray}
\sigma^2(P_{ij})=g_{i}^2g_{j}^2K_{i}K_{j}k^2 B^2 S_{C}^2  + g_{i}^2 g_{j}^2 k^{2}B^{2}T_{{\rm sys}i}T_{{\rm sys}j}
\label{eqn:sig2PijB}
\end{eqnarray}
We divide (\ref{eqn:sig2PijB}) by $g_{i}^2g_{j}^2K_{i}K_{j}k^2 B^2 S_{C}^2$, which is obtained from the foregoing calibration step. After averaging, the measured standard deviation of the output of the interferometer ($\Delta S_{ij}$) can be related to the expected value by
\begin{eqnarray}
\Delta S_{ij}&=&\frac{1}{\sqrt{2Bt_{\rm acc}}}\sqrt{S_{C}^2+\mathrm{SEFD}_{i}\mathrm{SEFD}_{j}}
\label{eqn:delSij}
\end{eqnarray}
with unit Jy and where $t_{\rm acc}$ is the averaging duration. As discussed in Sec.~\ref{sec:defn}, $S_{C}$ is negligible compared to the AAVS0.5/MWA $\mathrm{SEFD}$ for baselines $>\sim30\lambda$. Thus (\ref{eqn:delSij}) simplifies to
\begin{eqnarray}
\Delta S_{ij}\approx\sqrt{\frac{\mathrm{SEFD}_{i}\mathrm{SEFD}_{j}}{2Bt_{\rm acc}}}.
\label{eqn:delSijapprox2}
\end{eqnarray}
For $N$ baselines, we again obtain $N(N-1)/2$ pairs of such equations which can be solved for each $\mathrm{SEFD}$ through least-squares~\cite{Fen12}.

\section{Measurement Results}
\label{sec:meas}

\subsection{Calibration sources}
The measurement of  $A/T$ through cross-correlation requires a compact astronomical source that is sufficiently bright to obtain a good S/N calibration gain solution from a short ($\sim$minutes) scan. Thus the measurement of tile gain is restricted to the path in $(\theta, \phi)$ traced by calibrator sources moving across the beam of the AUT as the sky apparently rotates. 
Furthermore we seek to make measurements close to zenith, the direction of maximum gain. For a given source, the smallest zenith angle occurs at the meridian, a great circle passing through the zenith and the celestial poles (i.e. Az=0\degree or Az=180\degree). 

Tab.~\ref{tab:sources} lists the southern calibrator sources used for sensitivity measurement, selected to balance angular extent, brightness and zenith angle when transiting the meridian. They are not the brightest sources visible at the MRO, but other sources transit at larger zenith angles or are of considerable spatial-extent and difficult to model.
 
\begin{table} [htb]
\centering
\caption{Calibrator sources chosen for sensitivity measurement.}
\begin{tabular}{|l|p{2cm}|p{2.5cm}|}
\hline 
Source & ZA at meridian (deg.) & Indicative flux density at 160~MHz (Jy)\tabularnewline
\hline 
\hline 
3C444 & 9.7 & 80\tabularnewline
\hline 
Hydra A & 14.6 & 310\tabularnewline
\hline 
Hercules A & 31.7 & 520\tabularnewline
\hline 
\end{tabular}
\label{tab:sources}
\end{table}

\subsection{Calibration and imaging}
Calibration ensures coherence between all the independent visibility measurements in a 2-minute snapshot observation.
The accuracy of the antenna-based gain solutions from calibration ($g_{i}\sqrt{K_{i}}$) are normally qualitatively assessed. Their amplitude and phase should vary predictably with frequency, consistent with known instrumental effects such as differences in cable lengths.

Basic aperture synthesis imaging via the Fourier transform is another tool to verify calibration as it highlights errors in calibration that are not readily apparent in the visibility domain~\cite{Eke99}.
The MWA tiles and the AAVS0.5 forms a collection of 2-antenna cross-correlation interferometers (i.e. baselines) of various separation lengths to approximate a filled aperture; by extension of (\ref{eqn:visibility1D}), there is, to a good approximation, a 2-D Fourier relationship\footnote[10]{The Fourier relationship can be approximated as 2-D because the MWA is reasonably approximated as a co-planar array on snapshot integration timescales.} between the visibilities and the image of the incident radiation. 

\begin{figure} [htb]
	\begin{center}
	{\includegraphics[height=2.5in]{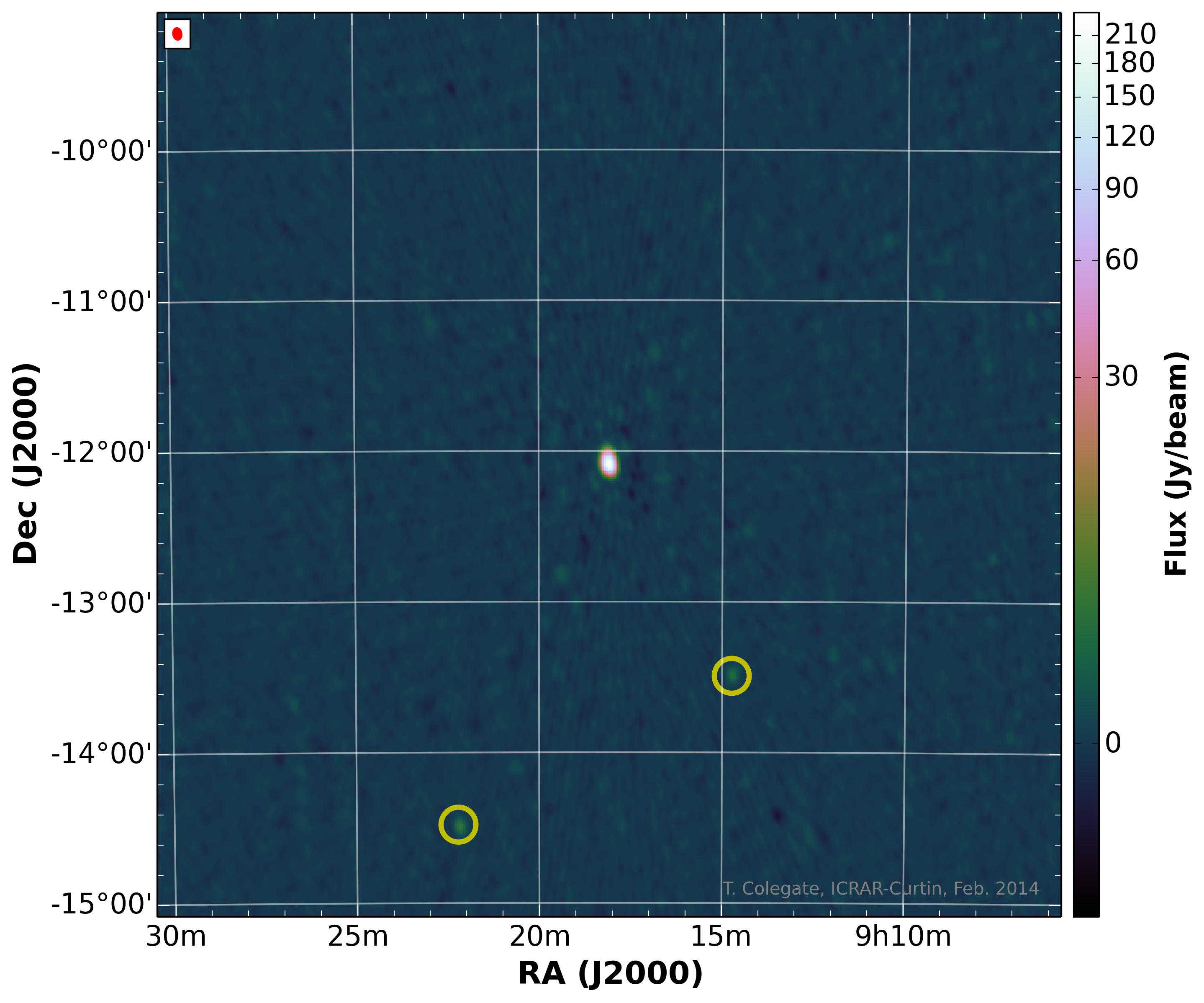}}
	
	{\includegraphics[height=2.5in]{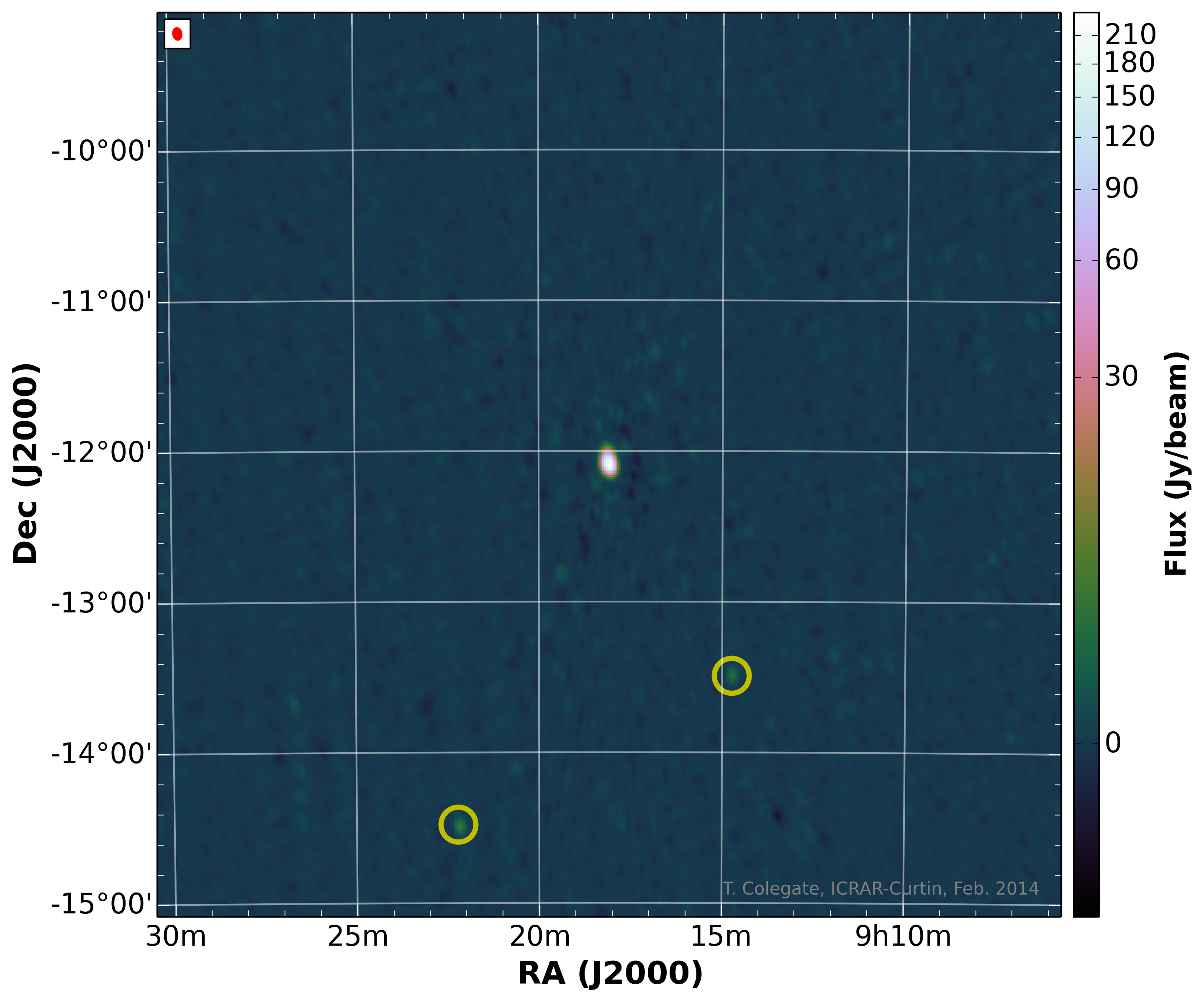}}
	\end{center}
\caption{Image of Hydra A (center) and the field out to 3\degree{} radius for X (top) and Y (bottom) antenna polarizations, centered at 155~MHz (30.72~MHz bandwidth) and using only MWA--AAVS0.5 baselines. Observation is 18 January 2014, 1:43 local time. The red oval (top-left) shows the size of the synthesis telescope beam. The yellow circles mark the positions of the next-strongest sources as recorded in the Molonglo Reference Catalogue of Radio Sources~\cite{LarMil81}.}
\label{fig:image-HydA}
\end{figure}

Fig.~\ref{fig:image-HydA} shows a dual-polarization image of the calibrator source Hydra~A.
Although the full set of MWA--MWA and MWA--AAVS0.5 baselines were used for calibration, only the 127 baselines involving the AAVS0.5 were used to create the image, thus verifying the AAVS0.5 functionality as an interferometer component. At this frequency, these baselines are not of sufficient length to resolve the spatial structure in Hydra~A. Other sources in the sky close (within $\sim 2\degree{}$) to Hydra A are significantly weaker but they align with known positions, demonstrating successful calibration. The X and Y-polarization images are similar, demonstrating consistency between the two signal chains.

\subsection{Sensitivity}
The measurement of the AAVS0.5 sensitivity was conducted with the MWA over several nights during 2014 and early 2015. The observations were made with the MWA tiles and the AAVS0.5 pointing at a zenith angle on the meridian such that the calibrator source passes through beam maximum. The MWA is limited to observing a total of 30.72~MHz of bandwidth at any one time. To sample the full 75--300~MHz frequency range of the MWA, we use two approaches: (i) stepping across the band over multiple nights and (ii) observing widely separated spot frequencies  during one night's observation.

\begin{figure} [htb]
    \begin{center}
    {\includegraphics[width=\columnwidth]{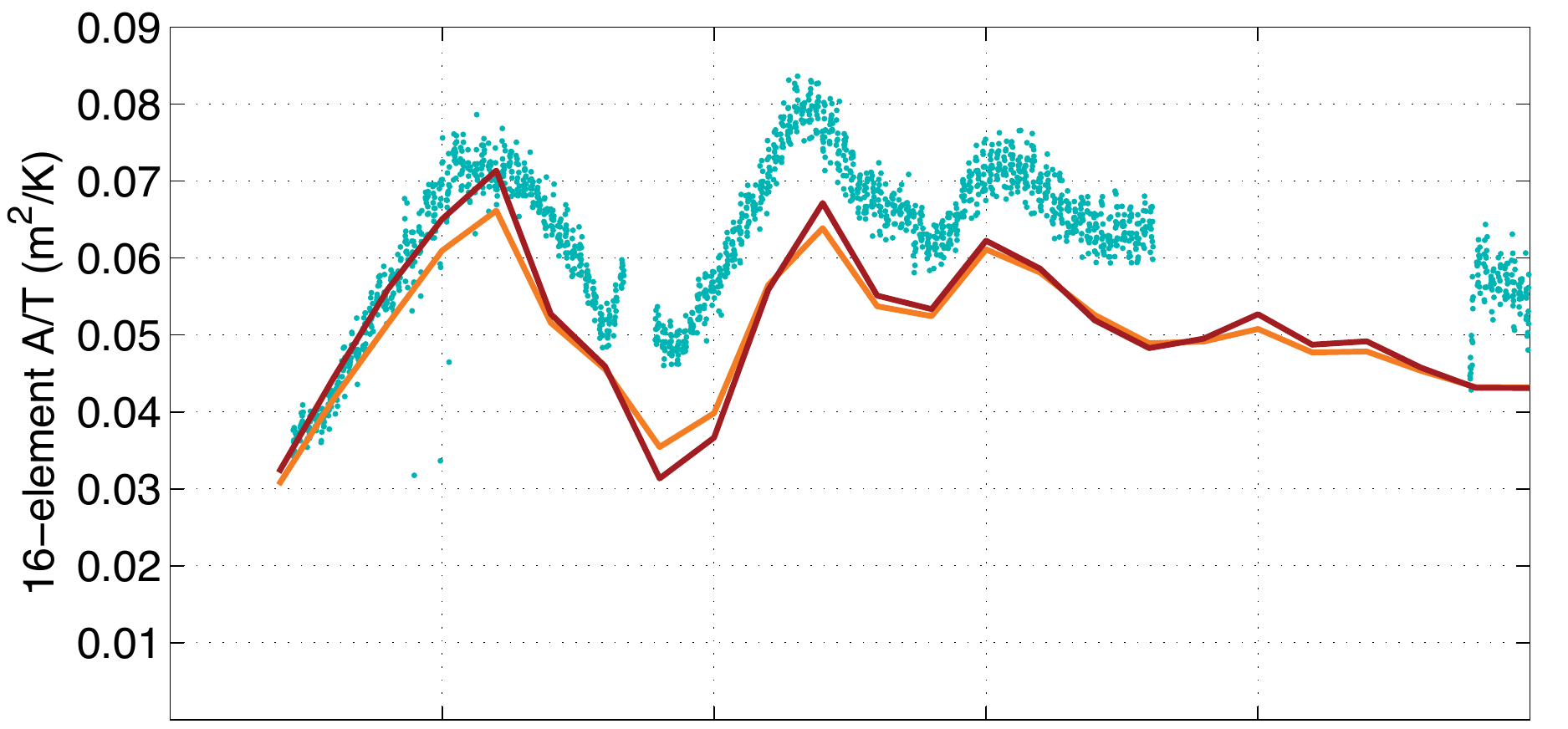}}
    {\includegraphics[width=\columnwidth]{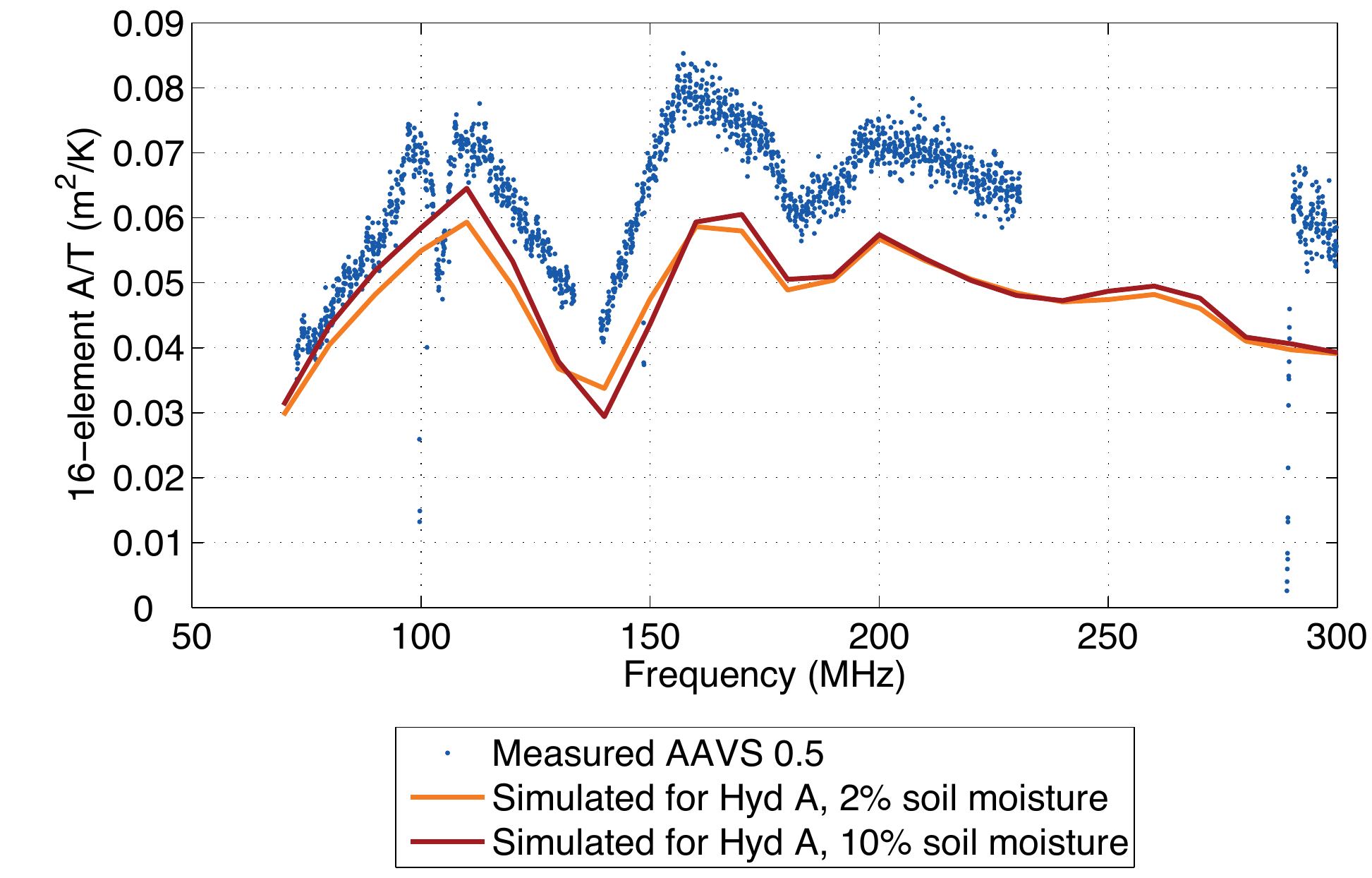}}
    \end{center}
\caption{Measured $A/T$ for a 2 minute observation of Hydra A starting at local sidereal time 9.27~h, taken over the period 14--21 January 2015, X~polarization (East--West arm) top, Y~polarization (North--South arm) bottom.  The AAVS0.5 pointing is Az=0\degree, ZA=14.6\degree.  Simulated $A/T$ also shown; see text for details.}
\label{fig:AonT-sweep}
\end{figure}

Fig.~\ref{fig:AonT-sweep} shows $A/T$ measurements made with Hydra A over the period 14--21 January 2015, for the X (E-W) and Y (N-S) antenna polarizations. 
The observing strategy involved 6$\times$2-minute snapshots of different 30.72~MHz  bands to cover the full MWA frequency range.
Observing each snapshot on a different night ensures the $A/ T$ measurement across the band is accurate to an angular resolution of 0.5\degree{} (the apparent motion of the calibrator source).
The lack of data at $\sim$138~MHz and $\sim$240--285~MHz is due to the persistent satellite-based RFI at these frequencies. 
Fig.~\ref{fig:AonT-sweep} also shows results expected from simulation for soil with 2\% and 10\% moisture (Sect.~\ref{sec:full-wave}). These represent a reasonable range in the likely soil moisture levels experienced at the MRO. 

The frequency-dependent trends in Fig.~\ref{fig:AonT-sweep} show good agreement between simulated and measured results. 
The degradation in $A/T$ between 100--150~MHz is due to the log-periodic antenna interaction with the ground. Design optimization to minimize this undulation is being explored.

\begin{figure} [tb]
    \begin{center}
    {\includegraphics[width=\columnwidth]{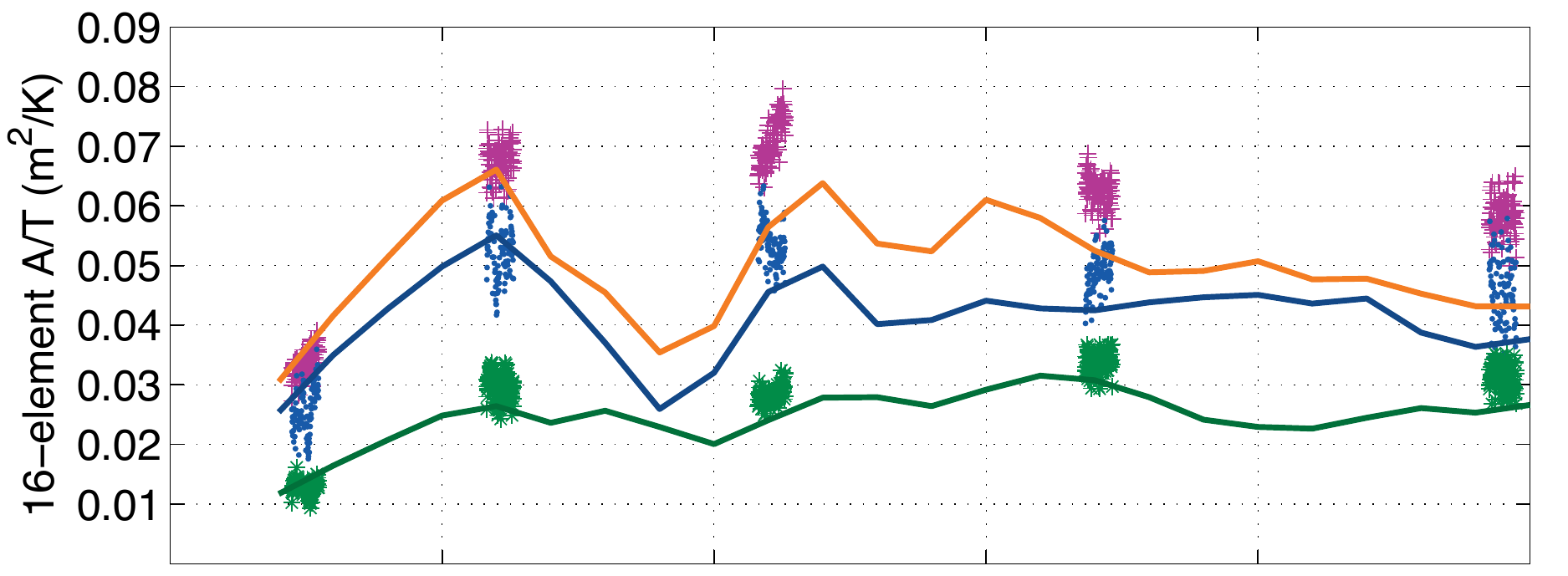}}
    {\includegraphics[width=\columnwidth]{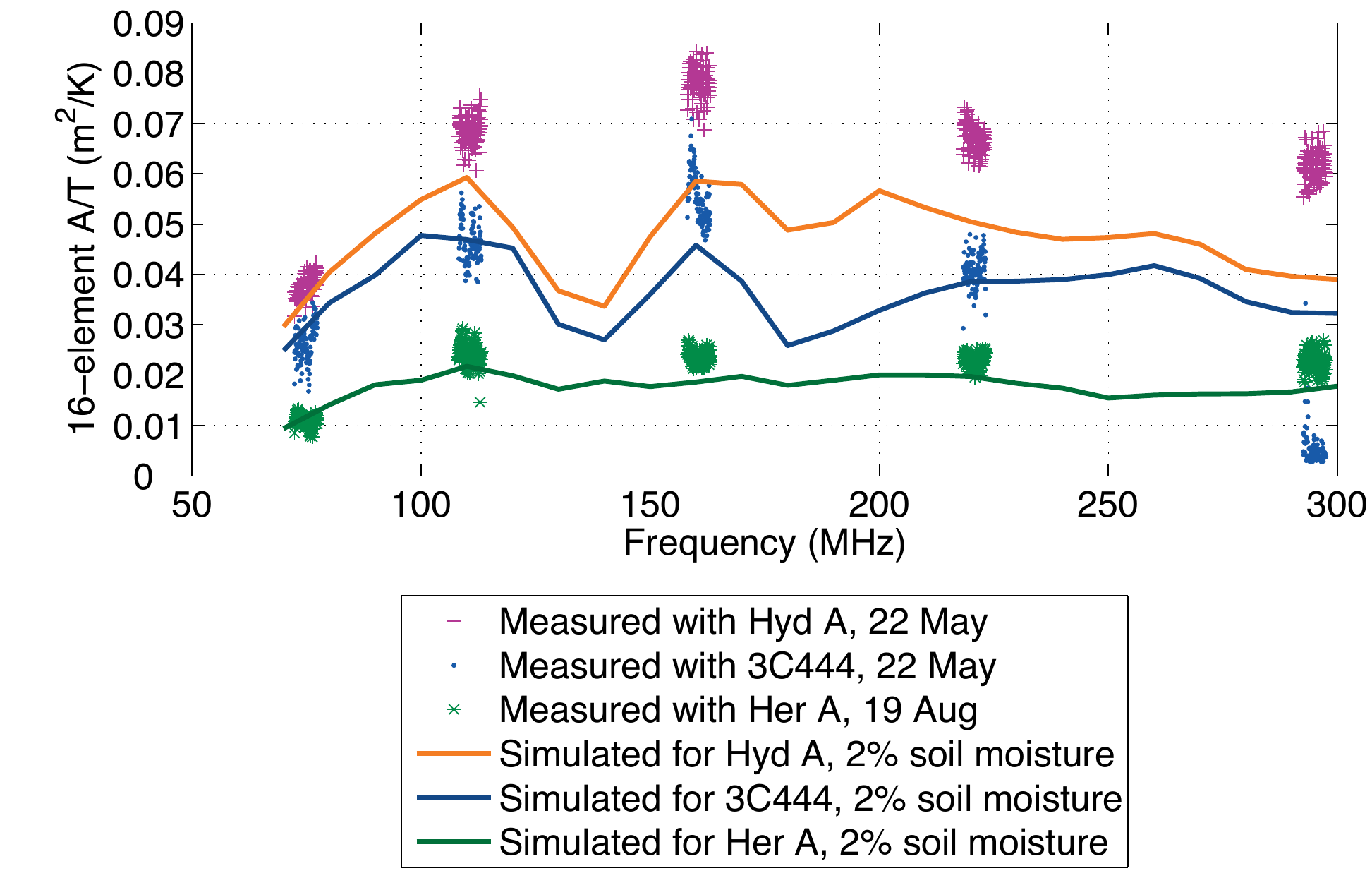}}
    \end{center}
\caption{Measured $A/T$ for the AAVS0.5 tile X polarization (top) and Y polarization (bottom), pointing at Az=0\degree{} and (top to bottom of each plot): ZA=14.6\degree{} (Hydra~A, 22 May 2014, 17:30:32 local time); ZA=9.7\degree{} (3C444, 23 May 2014, 6:24:00); and ZA=31.7\degree{} (Hercules~A, 19 August 2014, 19:12:24). Simulated $A/T$ for each pointing is also shown.}
\label{fig:AonT-3source}
\end{figure}

There is also a discrepancy between measurement and simulation at frequencies greater than $\sim$150~MHz, where measurement is consistently higher than simulation (up to $\sim$50\% at the higher frequencies).
Comparison with other calibrator sources indicates that this is likely related specifically to Hydra A.
 Fig.~\ref{fig:AonT-3source} plots $A/T$  measured for the AAVS0.5 beam pointed at three different calibrator sources on the meridian at zenith angles 9.7\degree~(3C444), 14.6\degree~(Hydra A) and 31.7\degree~(Hercules A), and the corresponding simulations for 2\% soil moisture.
For these data, the 30.72~MHz total bandwidth is split over the spot frequencies of interest in a single 2-minute snapshot to sample the full MWA frequency range in a single night. 
The measurement-to-simulation ratio ranges from approximately $-20$ to 20\% (3C444)\footnote[11]{The extremely small Y~polarization $A/T$ measured with 3C444 at 295~MHz arises from a corruption in the MWA calibration routine at this frequency and polarization, due to the presence of the Galactic plane in the MWA tile's grating lobe close to the western horizon. The X~polarization is not affected because the dipole beamshape makes that polarization much less sensitive at the eastern and western horizons.}  and $-10$ to 30\% (Hercules~A), indicating a discrepancy specific to the Hydra~A measurement.

Our hypothesis is that the error likely arises during the calibration procedure, due to incrorrect assumptions about the frequency-dependent flux density contribution from Hydra~A, i.e. $S_{\rm C}$ in (\ref{eqn:Pij_corr}).
Our models of the calibrator sources are derived from images from the Very Large Array Low-frequency Sky Survey Redux (VLSSr) at 74~MHz~\mbox{\cite{LanCot14}}, where the higher frequencies are modeled by a power-law scaling calculated from the total flux density measured at 74~MHz and 1.4~GHz~\mbox{\cite{ConCot98}}. An interpolated power-law scaling is necessary because there have been few well-calibrated measurements of these sources sampling the 70-300~MHz frequency range.

For Hydra~A, our model predicts the flux density at 300~MHz to be $\sim$1/3 that at 74 MHz. 
However, detailed measurements specifically targeting Hydra~A at 74 and 330~MHz~\mbox{\cite{1538-3881-127-1-48}} show that the flux density of the compact central region reduces less steeply with frequency, and would be only  $\sim$1/2 that at 74 MHz.  Unlike the surrounding diffuse emission, the compact central region is not spatially filtered at the higher frequencies, thus our modelled "known" flux density causes an underestimate of the actual flux density, resulting in a higher than actual $A/T$. In future processing of measurement, we expect significant convergence as the models of calibrator sources are improved for the MWA and other wide-band low-frequency telescopes.

\subsection{Beam pattern measurement}
\label{sub:beam-pattern}
\begin{figure} [htb]
    \begin{center}
     {\hfill\includegraphics[width=\columnwidth]{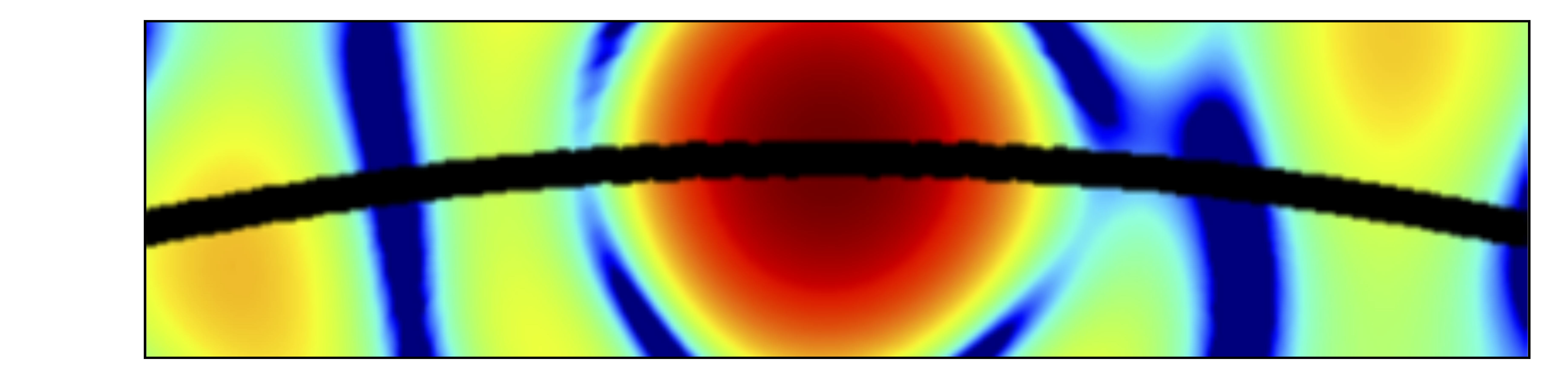}}
    {\includegraphics[width=\columnwidth]{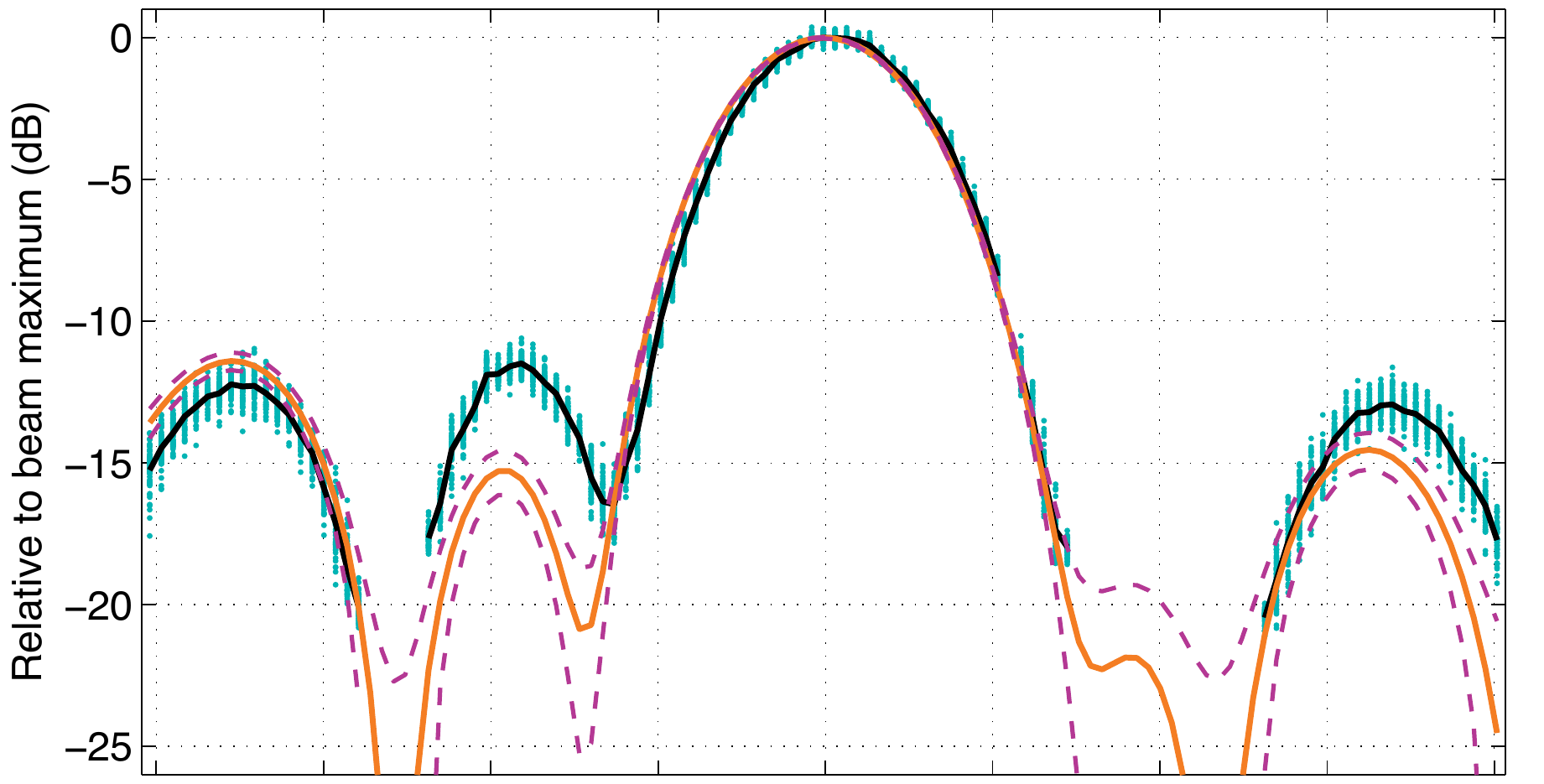}}
    {\includegraphics[width=\columnwidth]{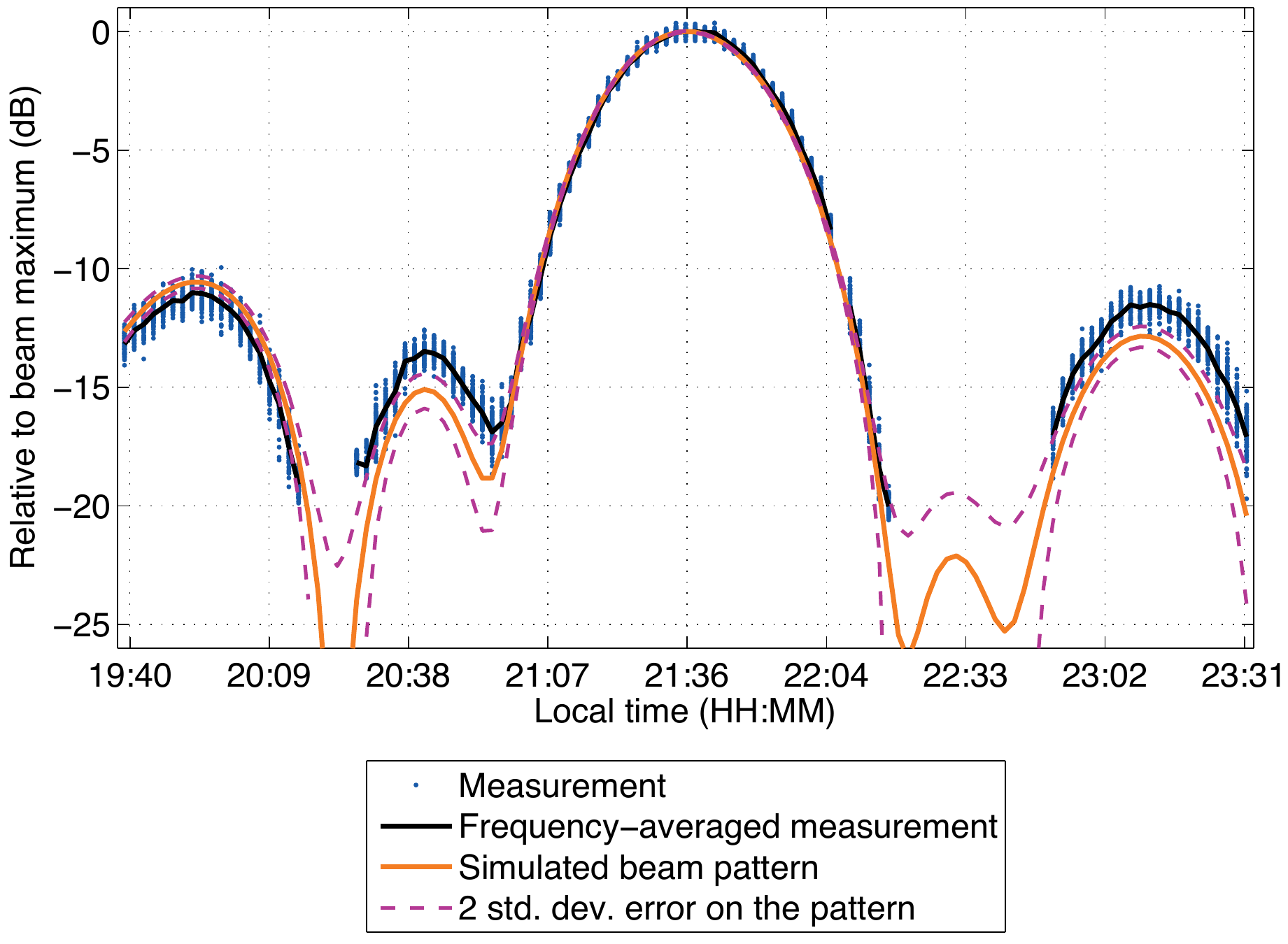}}
    \end{center}
\caption{The AAVS0.5 X (top and middle) and Y (bottom) polarization beam patterns at 220~MHz for Az=0\degree{}, ZA=14.6\degree{} pointing.
Hydra A was continuously observed on 21 March 2014 with 2-minute snapshots and 5.12~MHz bandwidth. 
Each data point is a 40~kHz channel measurement. The black curve is the same data, frequency-averaged for each snapshot. 
The orange curve shows the simulated (``error-free'') beam pattern for 2\% moisture. For clarity, the 10\% moisture case is not shown, as the difference is not significant.
The purple curve is the 2 standard deviation uncertainty on the simulated beam pattern. Each dataset is normalized individually. The inset at the top shows the trajectory of Hydra A (black curve) through the simulated X polarization beam pattern.}
\label{fig:beam-pattern}
\end{figure}

The sensitivity measurements are made at beam maximum by aligning the Az, ZA  direction of the AAVS0.5 beam pointing with the $\theta,\phi$ location of the calibrator source.
To characterize the beam pattern, 2-minute snapshot observations were taken for the same AAVS0.5 beam pointing, but over 4 hours prior to, during, and after source transit. For each snapshot, calibration is performed and $A/T$ calculated, providing a ``cut''  on the AUT's beam (fixed pointing direction) that coincides with the source's trajectory (tracked with the MWA). 
Fig.~\ref{fig:beam-pattern} shows the X and Y~polarization patterns measured with Hydra~A, and the cut through the simulated beam corresponding to the trajectory from $\theta$=31\degree{}, $\phi$=68\degree{} through the meridian at $\theta$=0\degree, $\phi$=14.6\degree to $\theta$=31\degree{}, $\phi$=293\degree{}.

In Fig.~\ref{fig:beam-pattern} the measured mainlobes are in very good agreement with simulation. The first sidelobe of the X-polarized pattern is within approximately 4~dB of simulated results and the others 2~dB or less.
Assuming random errors in the analog beamformer, the dashed purple curves show the estimated $\pm2$ standard deviation patterns, calculated per Appendix~\ref{app:beam-errors} for measured $\sigma_{\phi}=0.069$~radians in phase and $\sigma_{\rm A}=10\%$ in amplitude at this frequency. 
For the most part, we note that the $\pm2$ standard deviation patterns are consistent with the measured data including uncertainties. There remains, however, a $\sim$2 to 3~dB gap for the first sidelobe of the X-polarization which seems larger than could be explained by random errors alone. This would suggest a systematic error, the cause of which we are still investigating.

Beam pattern measurement will likely be an important tool to verify SKA requirements for a polarization and frequency-dependent beam model, described as a function of $\theta, \phi$~\cite{SKA14-SRS-L1-Rev4}. 
Investigation of computationally efficient beam models is ongoing\cite{LerRaz11-mutual,LerCra13,SutOSu15}; Fig.~\ref{fig:beam-pattern} shows that we can provide partial verification of such models, noting that the specific cuts through the beam pattern available for measurement depend on the beam pointing direction and the trajectory of available calibration sources.
The lower cut-off of the pattern measurement is approximately $-18$~dB in Fig.~\ref{fig:beam-pattern} and $-22$~dB at 110~MHz (not shown); this cut-off is where there is insufficient cross-correlated S/N ($S_{ij}/\Delta S_{ij}$) on baselines involving AAVS0.5. In comparison, $-25$~dB is achieved with a UAV~\cite{6704306} and $-30$~dB with  an opportunistic satellite-based measurement technique at 137~MHz~\cite{NebBra15}. 

More fundamentally for our measurements, the limit to dynamic range depends on the contributors to S/N ratio:  the brightness of the calibrator source and sensitivity of the antenna pairs. 
Assuming the sensitivity of the other antenna in the pair remains constant, ({\ref{eqn:delSijapprox2}) shows that the S/N ratio scales with the square root of the AUT sensitivity at $\theta,\phi$.
With the AAVS1 array $\sim$16 times more sensitive than the AAVS0.5, we expect an additional 6~dB dynamic range, and thus the main lobe and sidelobes can be measured to $-24$ to $-28$~dB.

\section{Conclusion}
\label{sec:concl}
We have further demonstrated and validated the AAVS0.5 as a highly capable prototype for low-frequency aperture arrays. Using the AAVS0.5 in cross-correlation mode with the MWA interferometer telescope, we have made sensitivity and beam pattern measurements that generally show good agreement with simulation. Where there are differences between measurement and simulation, we have identified a path to achieve convergence through improved calibrator source models. Our results demonstrate that interferometric measurement is highly useful in evaluating array sensitivity and sidelobe performance. Furthermore, electromagnetic simulation, properly employed, can be used to accurately estimate array performance.

This exercise of fully utilizing the AAVS0.5 system serves as a good template for the design, deployment and operation of pre-construction engineering prototypes for the low-frequency SKA. We plan to apply the methods discussed here to the envisaged AAVS1 array, the next-generation SKA station-sized low-frequency aperture array. We will again cross-correlate the AAVS1 signals with the MWA and process the data in a manner similar to the AAVS0.5. 
With the factor of $\sim$16 increase in the AAVS1 sensitivity compared to the AAVS0.5, we expect to achieve a factor of 4 reduction in measurement uncertainty and also a similar improvement in pattern measurement dynamic range. The increased sensitivity will also enable $A/T$  measurement via fainter and more point-like sources close to zenith. Characterization through astronomical measurement is enhanced by the more sensitive array, and presents a clear path forward for sensitivity and beam pattern characterization with results that are directly relevant to the low-frequency SKA.

\appendices

\section{Beam pattern errors}
\label{app:beam-errors}

Using array tolerance theory, we estimate the error from ideal beam due to beamformer and RF component tolerances. 
We apply the formulas found in classic references~\cite{Zucker_ant1_1969_ch6, Mailloux_2005_ch7,Hansen_1998_ch7} assuming uniform amplitude excitation and small random errors.

We obtained an estimate of the total combined variance due to phase and amplitude errors through measurement of the system shown in Fig.~\ref{fig:AAVS_RX} from the LNA to the MWA beamformer. From a population of 16 RF chains, at 220~MHz, we estimate
\begin{align}
\sigma^{2} & =\sigma_{\phi}^{2}+\sigma_{A}^{2},\\
 & =0.07^{2}+0.1^{2}=0.0148 \nonumber
\end{align}

The ensemble mean beam pattern that incorporates these errors is~\cite{Zucker_ant1_1969_ch6, Mailloux_2005_ch7}
\begin{equation}
\lvert\overline{f_{n}(\theta,\phi)}\rvert^{2}=\lvert f_{n0}(\theta,\phi)\rvert^{2}+\frac{\sigma^{2}}{N},
\label{eqn:mean_pat}
\end{equation}
where $\lvert f_{n0}(\theta,\phi)\rvert^{2}$ is the normalized error-free
power pattern and $N$ is the number of elements. The range of ZA shown in Fig.~\ref{fig:beam-pattern} is close to the main beam such that the ensemble mean is negligibly different to the error-free pattern since $\lvert f_{n0}(\theta,\phi)\rvert^{2}>>\sigma^{2}/N$. The statistics of the field pattern for $\lvert f_{n0}(\theta,\phi)\rvert^{2}>>\sigma^{2}/N$ approaches a Gaussian distribution with variance $\sigma^{2}/(2N)$~\cite{Zucker_ant1_1969_ch6,Mailloux_2005_ch7}. Hence the one standard deviation field pattern may be written as
\begin{equation}
\lvert f_{n0}(\theta,\phi)\rvert\pm\frac{\sigma}{\sqrt{2N}}.
\end{equation}

\section*{Acknowledgment}
We acknowledge the Wajarri Yamatji people as the traditional owners of the Murchison Radio Astronomy Observatory site. The MRO is operated by CSIRO, whose assistance we acknowledge.  The AAVS0.5 operates as an external instrument of the MWA telescope and we thank the MWA project and personnel, in particular, D. Emrich, B. Crosse and A. Williams for their support. We also acknowledge B. Fiorelli of ASTRON and F. Schlagenhaufer of Curtin University for their contributions to the implementation of the AAVS0.5 and
discussions surrounding array metrology.


\end{document}